\documentclass[12pt]{article} 

\usepackage{latexsym} 
\usepackage{amssymb}  
\usepackage{epsfig}       

\newcommand{\beq}{\begin{equation}}
\newcommand{\eeq}{\end{equation}}
\newcommand{\ba}{\begin{array}}
\newcommand{\bea}{\begin{eqnarray}}
\newcommand{\ea}{\end{array}}
\newcommand{\eea}{\end{eqnarray}}

\renewcommand{\slash}{\!\!\!\!/\,}

\newcommand\comment[1]{ \hbox{[{\it Comment suppressed here.}\/]} }
\newcommand\hide[1]{}

\newcommand{\tr}{\hbox{tr}}
\newcommand{\Tr}{\hbox{Tr}}

\newcommand{\bx}{{\bf x}}
\newcommand{\by}{{\bf y}}

\newcommand{\bp}{{\bf p}}
\newcommand{\bq}{{\bf q}}
\newcommand{\bk}{{\bf k}}

\newcommand{\ap}{|{\bf p}|}
\newcommand{\aq}{|{\bf q}|}
\newcommand{\ak}{|{\bf k}|}

\newcommand{\skipover}[1]{}
\newcommand{\nn}{\nonumber \\}

\newcommand{\C}{\mathcal{C}}

\makeatletter 

\def\appendix{\par                              
    \setcounter{section}{0}                     
    \setcounter{subsection}{0}
    \renewcommand{\theequation}{\Alph{section}.\arabic{equation}}
    \renewcommand{\thesection}{Appendix \Alph{section}}
}

\def\applabel#1{\@bsphack
  \protected@write\@auxout{}%
         {\string\newlabel{#1}{{\Alph{section}}{\thepage}}}%
  \@esphack}

\def\section{
\setcounter{equation}{0}        
\@startsection {section}{1}{\z@}{-3.5ex plus -1ex minus 
 -.2ex}{2.3ex plus .2ex}{\large\bf}}
\renewcommand{\theequation}{\arabic{section}.\arabic{equation}}

\def\subsection{\@startsection{subsection}{2}{\z@}{-3.25ex plus -1ex minus 
 -.2ex}{1.5ex plus .2ex}{\normalsize\bf}}

\def\subsubsection{\@startsection{subsubsection}{3}{\z@}{-3.25ex plus
 -1ex minus -.2ex}{1.5ex plus .2ex}{\normalsize}}

\makeatother   

\newsavebox{\eqlabel}

\makeatletter  
\newlength{\numblen}
\newsavebox{\eqnumb}
\def\@eqnnum{\savebox{\eqnumb}{\rm (\theequation)}%
\settowidth{\numblen}{\usebox{\eqnumb}}%
\makebox[\numblen][l]{\usebox{\eqnumb}~~~\usebox{\eqlabel}}}
\makeatother   

\newenvironment{equationwithlabel}[1]{ %
  \savebox{\eqlabel}{#1}
  \begin{equation}\label{#1} }{\end{equation}} 
\newcommand{\beql}[1]{\begin{equationwithlabel}{#1}}
\newcommand{\eeql}{\end{equationwithlabel}}

\begin{document}

\title{\bf $n$-Particle irreducible 
effective action techniques for gauge theories
\\[1.ex]}

\author{
J\"urgen Berges\thanks{email: j.berges@thphys.uni-heidelberg.de}\\ 
\normalsize{Universit\"at Heidelberg, Institut f\"ur 
Theoretische Physik}\\
\normalsize{Philosophenweg 16, 69120 Heidelberg, Germany}
}

\date{}

\begin{titlepage}
\maketitle
\def\thepage{}          

\begin{abstract}
\noindent
A loop or coupling expansion of a so-called $n$-particle irreducible
($n$PI) generating functional provides a well-defined approximation 
scheme in terms of self-consistently dressed propagators and 
$n$-point vertices. A self-consistently complete description determines 
the functional for arbitrarily high $n$ to a given order in the 
expansion. We point out an equivalence hierarchy for $n$PI effective 
actions, which allows one to obtain a self-consistently complete result
in practice. The method is applied to a $SU(N)$ gauge theory with 
fermions up to four-loop or ${\mathcal O} (g^6)$ corrections. 
For non-equilibrium we discuss the connection to kinetic theory. 
The leading-order \mbox{on-shell} results in $g$ can be obtained 
from the three-loop effective action
approximation, which already includes in particular all diagrams enhanced 
by the Landau Pomeranchuk Migdal effect. Furthermore, we compare the effective
action approach with Schwinger-Dyson (SD) equations.  
By construction, SD equations are expressed in terms of loop diagrams 
including both classical and dressed vertices, which leads to 
ambiguities of whether classical or dressed ones should be 
employed at a given truncation order. We point out that these 
problems are absent using effective action techniques.    
We show that a wide class of truncations of SD 
equations cannot be obtained from the $n$PI effective action. 
In turn, our results can be used to resolve 
SD ambiguities of whether classical or dressed vertices should be employed 
at a given truncation order.
\end{abstract}

\end{titlepage}

\renewcommand{\thepage}{\arabic{page}}


\section{Introduction and overview}
\label{sec:intro}

\subsection{Background}
\label{sec:background}

Selective summation to infinite order in perturbation theory
often plays an important role in vacuum, thermal equilibrium
or nonequilibrium quantum field theory.
A prominent example in high temperature field theory 
is the so-called ``hard-thermal-loop'' (HTL) perturbation 
theory~\cite{Braaten:1989mz}.
For small coupling $g \ll 1$ the description of gauge boson 
excitations with wave number $k \sim g T$ requires 
appropriately resummed propagators and vertices. After the
selective HTL resummation the effective interactions among the
$g T$ scale degrees of freedom are weak and may be treated 
perturbatively. However, for excitations with wave number 
$k \sim g^2 T$ the occupation numbers of individual
modes can grow nonperturbatively large $\sim 1/g^2$ and 
the perturbative treatment breaks down.

For out-of-equilibrium situations there are additional 
complications which do not appear in vacuum or thermal 
equilibrium\footnote{This does not concern restrictions to
sufficiently small deviations from thermal equilibrium, such as described
in terms of (non-)linear response theory, which only involve
thermal equilibrium correlators in real time.}. 
Nonequilibrium dynamics typically poses 
an initial value problem: time-translation invariance is explicitly 
broken by the presence of the initial time, where the system has 
been prepared. During the nonequilibrium evolution the system may 
effectively loose the dependence on the details of the initial condition,
and become approximately time-translation invariant
for sufficiently late times. If thermal equilibrium is approached then the 
late-time result is universal in the sense that it becomes uniquely 
determined by the values of
the (conserved) energy density and of possible conserved
charges\footnote{Here we consider closed systems without 
coupling to a heat bath or external fields, which could provide 
sources or sinks of energy.}. 
It is well-known that the late-time
behavior of quantum fields cannot be described using standard perturbation 
theory. The perturbative time evolution suffers from
the presence of spurious, so-called secular terms, which grow with 
time and invalidate the expansion even in the presence of a weak 
coupling. Here it is important  
to note that the very same problem can appear as well for nonperturbative
approximation schemes such as $1/N$ 
expansions~\cite{Ryzhov:2000fy}.\footnote{Note that restrictions 
to mean-field-type approximations are insufficient. 
They typically suffer from the presence of an infinite number of spurious 
conserved quantities, and are known to fail to describe
thermalization.}\\ 

It has recently been demonstrated for 
scalar~\cite{Berges:2000ur,Berges:2001fi,Aarts:2001yn,Cooper:2002qd,Aarts:2001qa,Bedingham:2003hu} and 
fermionic~\cite{Berges:2002wr} theories that nonequilibrium
dynamics with subsequent late-time thermalization
can be described from a selective summation of powers of the 
coupling or $1/N$ without further assumptions. This provides 
an efficient solution to the problem of a universal late-time behavior 
as well as the secularity problem. These approximations are expressed 
in terms of a loop~\cite{Cornwall:1974vz} or 
$1/N$~\cite{Berges:2001fi,Aarts:2002dj} expansion of the so-called 
{\em two-particle irreducible} (2PI) effective 
action~\cite{Baym}.\footnote{Loop 
approximations of the 2PI effective
action are also called ``$\Phi$-derivable''.} 
Though other resummations may be
invoked to circumvent secular behavior of perturbative 
treatments~(cf.~e.g.~\cite{Boyanovsky:2003ui}), 
the description of a universal late-time 
behavior poses rather strong restrictions on the possible 
approximations. The 2PI schemes seem to be uniquely
suitable in nonequilibrium quantum field theory
to capture the effective loss of initial condition details leading
to thermalization.\footnote{Other 
approaches include truncated hierarchies for equal-time
correlators~\cite{Aarts:2001wi} 
or so-called ``two-point-particle irreducible''
schemes~\cite{Baacke:2003qh}, for which thermalization could not be 
demonstrated so far.} The remarkably good convergence properties of the 
approach have also been observed in the context of classical statistical 
field theories, where comparisons with exact results are 
possible~\cite{Aarts:2001yn}.
The expansions do not rely on small departures from equilibrium, or 
sufficient space-time homogeneity of the system underlying
effective kinetic descriptions in terms of 
``quasiparticles''~\cite{Blaizot:2001nr}. 
However, 2PI effective action techniques can be very efficient
in deriving kinetic equations~\cite{Calzetta:1988cq,Danielewicz,Ivanov:1998nv,Blaizot:2001nr,Prokopec:2003pj}. 

The 2PI expansions 
are known to be ``conserving''~\cite{Baym,Knoll:2001jx},
i.e.~they are consistent with global symmetries 
of the Lagrangian~\cite{vanHees:2002bv}.
In particular, energy conservation and the absence of an irreversible 
dynamics are viable ingredients for a description of 
nonequilibrium time evolution from first principles.   
However, these approximations can violate Ward identities 
associated to local 
symmetries, which has recently been explored and shown to be suppressed
with respect to naive estimates based on power 
counting~\cite{Arrizabalaga:2002hn}.
First applications in gauge theories use the
2PI effective action as an efficient starting point for the
development of selective summation 
schemes for the description of the equilibrium thermodynamics of
the quark-gluon plasma~\cite{Blaizot:1999ip,Peshier:2000hx} (cf.~also
\cite{BP}).

\subsection{Equivalence hierarchy for $n$PI effective actions}
\label{sec:npieffectiveactions}

To understand the success and, more importantly, the limitations 
of expansion schemes based on the 2PI effective action we 
consider in this paper so-called ``$n$PI'' effective actions
for $n > 2$. Recall that the description of the 2PI effective 
action $\Gamma[\phi,D]$ employs a self-consistently dressed 
one-point function $\phi$ and two-point function $D$: 
The field expectation value 
$\phi = \langle \varphi \rangle$ and connected propagator
$D = \langle T \varphi \varphi \rangle - \phi \phi$ are 
dressed by solving the equations of motion 
$\delta \Gamma/\delta \phi = 0$ and $\delta \Gamma/\delta D = 0$
for a given order in the (e.g.~loop) expansion of 
$\Gamma[\phi,D]$~\cite{Cornwall:1974vz}. 
However, the 2PI effective action does not treat the higher 
$n$-point functions with $n>2$ on the same footing as the lower ones:
The three- and four-point function etc.~are not self-consistently 
dressed in general, i.e.~the corresponding proper three-vertex 
$V_3$ and four-vertex $V_4$ are given by the classical ones. In contrast, 
the $n$PI effective action $\Gamma[\phi,D,V_3,V_4,\ldots,V_n]$ 
provides a dressed description 
for the proper vertices $V_3,V_4,\ldots,V_n$ as well,
with $\delta\Gamma/\delta V_3 =0,\delta\Gamma/\delta V_4 =0, 
\ldots ,\delta\Gamma/\delta V_n =0$.

For applications it can be desirable to obtain a self-consistently 
complete description, which to a given order in the expansion determines 
$\Gamma[\phi,D,V_3,V_4,\ldots,V_n]$ for arbitrarily high $n$.
Despite the complexity of a general $n$PI effective action
it is important to note that a systematic, e.g.~loop
or coupling expansion can be nevertheless performed in practice.
We point out that a self-consistently complete loop-expansion
of the effective action can be based on the following
equivalence hierarchy:
\bea
\Gamma^{\rm (1loop)}[\phi] &\!=\!& \Gamma^{\rm (1loop)}[\phi,D] =\ldots\, ,
\nonumber\\
\Gamma^{\rm (2loop)}[\phi] &\!\not =\!& \Gamma^{\rm (2loop)}[\phi,D]
= \Gamma^{\rm (2loop)}[\phi,D,V_3]=\ldots \, ,
\label{eq:hierarchy}\\
\Gamma^{\rm (3loop)}[\phi] &\!\not =\!& \Gamma^{\rm (3loop)}[\phi,D]
\not = \Gamma^{\rm (3loop)}[\phi,D,V_3] 
= \Gamma^{\rm (3loop)}[\phi,D,V_3,V_4]=\ldots \, ,
\nonumber
\eea   
where $\Gamma^{{\rm (}n{\rm -loop)}}$ denotes the approximation of the 
respective effective action to $n$-th loop order in the
absence of sources. As a consequence, for a theory as e.g.~quantum 
electrodynamics (QED) or chromodynamics (QCD) the 2PI
effective action provides a self-consistently complete
description to two-loop order or\footnote{Here, 
and throughout the paper, $g$ means the strong gauge coupling $g_s$ for 
QCD, while it should be understood as the electric charge $e$ for QED.
For the power counting we take $\phi \sim {\mathcal O} (1/g)$
(cf.~Sec.~\ref{sec:compu}). The metric is denoted as 
$g^{\mu\nu} = g_{\mu\nu} = \mbox{diag}(1,-1,-1,-1)$. 
} 
${\mathcal O}(g^2)$: For a two-loop approximation
all $n$PI descriptions with $n \ge 2$ are equivalent and
the 2PI effective action captures already the 
complete answer for the self-consistent description
up to this order. In contrast, a self-consistently complete 
result to three-loop order or ${\mathcal O}(g^4)$
requires at least the 3PI effective action etc.  
This hierarchy clarifies a number of questions in the 
literature about the success or insufficiency of expansion
schemes based on the 2PI effective action:

i) Recently, it was argued that for
high temperature gauge theories 
a loop-expansion of the 2PI effective action 
is not suitable for a quantitative 
description of transport coefficients in the context
of kinetic theory~\cite{Moore:2002mt}. 
As an example, the calculation of shear viscosity in a 
theory like QCD may be based on the inclusion of an infinite series of 
2PI ``ladder'' diagrams in order to recover the leading order 
``on-shell'' results in $g$~\cite{Arnold:2002zm}. 
The enhancement of the infinite series 
of apparently higher order diagrams can be understood as a manifestation 
of the Landau Pomeranchuk Migdal (LPM) effect~\cite{Aurenche:2000gf}.
As pointed out above, for gauge theories such as QCD the 2PI effective 
action provides a self-consistently complete description to two-loop 
order or ${\mathcal O}(g^2)$. However, to go beyond that order  
in this scheme requires to consider higher effective actions.
4PI effective actions for scalar field theories have been derived
previously in Ref.~\cite{4PI}. In Ref.~\cite{FMcL} the thermodynamic potential 
for QED with a full three-vertex was constructed, and 
a perturbative construction scheme for the 4PI effective action 
in QCD was given. We derive the 4PI effective
action for a nonabelian $SU(N)$ gauge theory with fermions
up to four-loop or ${\mathcal O} (g^6)$ corrections,
starting from the 2PI effective action and doing subsequent
Legendre transforms   
(Sects.~\ref{sec:highereffective actions},\ref{sec:nonabgauge}). 
The class of models include gauge 
theories such as QCD or abelian theories as QED, as well as simple 
scalar field theories with cubic or quartic interactions. 
For non-equilibrium (Sec.~\ref{sec:nonequilibrium})
we discuss the connection to kinetic theory in 
QED (Sec.~\ref{sec:kinetic}).
We will see that, since the lowest order contribution to the kinetic equation
is of ${\mathcal O}(g^4)$, the 3PI effective action provides 
a self-consistently complete starting point for its description.
In particular, the leading-order on-shell result in $g$ can be 
efficiently obtained from the 3PI effective action to three-loops,
which includes in particular all diagrams enhanced by the LPM effect. 

ii) In nonequilibrium quantum field theory the success of the
2PI effective action to describe a universal late-time behavior
(cf.~Sec.~\ref{sec:background}) crucially depends on the self-consistent
nature of the employed approximation scheme. We note that the
successful descriptions of thermalization in scalar~\cite{Berges:2000ur} 
and fermionic theories~\cite{Berges:2002wr} based on
a 2PI loop expansion were self-consistently complete in the
above sense: 
We show in Sec.~\ref{sec:highereffective actions} that 
in the absence of a three-vertex and spontaneous symmetry breaking,
to three-loop order the 2PI effective action already contains the 
complete answer for the self-consistent description up to this order:
$\Gamma^{\rm (3loop)}[\phi=0,D] = \Gamma^{\rm (3loop)}[\phi=0,D,V_3=0,V_4]$.
In the presence of (effective) cubic interactions the three-vertex
would receive further corrections from the 3PI effective action. 

iii) The evolution equations, which are obtained by variation of 
the \mbox{$n$PI} effective action, are closely related to 
Schwinger-Dyson (SD) equations~\cite{SD}. Without approximations 
the equations of motion obtained from an exact \mbox{$n$PI} 
effective action and the exact SD equations have to agree since one 
can always map identities onto each other. However, in general
this is no longer the case for a given order in the
loop or coupling expansion of the $n$PI effective action.
By construction, SD equations are expressed in terms of loop diagrams 
including both classical and dressed vertices, which leads to 
ambiguities of whether classical or dressed ones should be 
employed at a given truncation order.     
In particular, SD equations are not closed a priori in the sense that
the equation for a given $n$-point function always involves
information about $m$-point functions with $m > n$. 

We point out that these problems are absent using effective action 
techniques (Sec.~\ref{sec:equationsofm}). 
In turn, we show that a wide class of truncations of exact SD 
equations cannot be obtained from the $n$PI effective action. 
In particular, our results can be used to resolve 
ambiguities of whether classical or dressed vertices should be employed 
for a given truncation of a SD equation. For instance, in QCD
the three-loop effective action result leads to evolution equations, which 
are equivalent to the SD equation for the two-point 
function and the one-loop three-point function {\em if}$\,$ all
vertices in loop-diagrams for the latter are replaced by the full 
vertices at that order\footnote{Disagreements 
of recent results in scalar $\phi^4$-theory inferred 
from the three-loop 4PI effective action as compared to earlier
results~\cite{4PI} are due to additional 
approximations for the
vertices in Ref.~\cite{Carrington:2004sn}.}  
As mentioned above the ``conserving'' property 
of using an effective action truncation can have important 
advantages, in particular if 
applied to nonequilibrium time evolution problems, where the 
presence of basic constants of motion such as energy conservation 
is crucial. SD equations have been frequently applied to nonperturbative 
strong interaction physics and, for instance,
recent comparisons of certain approximations with gauge-fixed 
lattice results are encouraging~\cite{Alkofer:2000wg}, 
also for effective action techniques.\\

\section{Higher effective actions}
\label{sec:highereffective actions}

All information about  
the quantum theory can be obtained from the effective 
action, which is a generating functional
for Green's functions. Typically, the (1PI) effective action 
$\Gamma[\phi]$ is represented as a functional of the --- bosonic or
fermionic --- field expectation 
value or one-point function $\phi$ only. In contrast, the so-called 
2PI effective action $\Gamma[\phi,D]$ is conventionally written as a functional
of $\phi$ and the full propagator or connected two-point 
function $D$~\cite{Baym,Cornwall:1974vz}. 
The latter provides an efficient description
of quantum corrections in terms of loop-diagrams with dressed propagator
lines and classical vertices. The functional 
dependence of higher effective actions take into account as well 
the dressed three-point function, four-point function etc.~or, 
equivalently, the proper three-vertex
$V_3$, four-vertex $V_4$ and so on~\cite{4PI,Calzetta:1988cq}. 
The name ``3PI'' effective action is 
used to denote $\Gamma[\phi,D,V_3]$, and ``4PI'' refers to 
$\Gamma[\phi,D,V_3,V_4]$ and similarly for even higher
effective actions. The functionals are constructed 
such that the equations of motion for the respective ``field variables'' 
are obtained from stationarity conditions:
\bea
\frac{\delta \Gamma[\phi]}{\delta \phi} = 0
\eea 
for the 1PI effective action, and
\bea
\frac{\delta \Gamma[\phi,D]}{\delta \phi} = 0
\quad , \quad 
\frac{\delta \Gamma[\phi,D]}{\delta D} = 0
\label{eq:2PIeqnmot}
\eea 
for the 2PI action in the absence of sources, etc.

All functional representations of the
effective action are equivalent in the sense that they are
generating functionals for Green's functions including all 
quantum/statistical fluctuations and, in the absence of
sources, have to agree by construction:
\bea
\Gamma[\phi] = \Gamma[\phi,D] = \Gamma[\phi,D,V_3] 
= \Gamma[\phi,D,V_3,V_4]
= \Gamma[\phi,D,V_3,V_4,\ldots,V_n] 
\eea  
for arbitrary $n$ without further approximations. However, e.g.~loop expansions
of the 1PI effective action to a given order in the presence of 
the ``background'' field $\phi$ differ in general from a loop expansion
of $\Gamma[\phi,D]$ in the presence of $\phi$ and $D$. A similar
statement can be made for expansions of higher functional
integrals. For a $n$PI effective action at a given expansion order 
all $\phi$, $D$, $V_3$, \ldots, $V_n$ are self-consistently
determined by the stationarity conditions similar to 
(\ref{eq:2PIeqnmot}). As mentioned in the introduction,
for applications it is often desirable to obtain a self-consistently 
complete description, which to a given order in the expansion determines 
$\Gamma[\phi,D,V_3,V_4,\ldots,V_n]$ for arbitrarily high $n$. 
For practical purposes it is important to realize that there 
exists an equivalence hierarchy 
as displayed in Eq.~(\ref{eq:hierarchy}) such that feasible calculations 
with lower effective actions are sufficient. As is
shown in Sec.~\ref{sec:selfcomplete}, 
for instance at three-loop order one has:
\bea
\Gamma^{\rm (3loop)}[\phi] \not = \Gamma^{\rm (3loop)}[\phi,D]
\not = \Gamma^{\rm (3loop)}[\phi,D,V_3] 
&\! =\!& \Gamma^{\rm (3loop)}[\phi,D,V_3,V_4] \\ 
&\! =\!& \Gamma^{\rm (3loop)}[\phi,D,V_3,V_4, \ldots,V_n]\, , \nonumber 
\eea
to arbitrary $n$ in absence of sources. As a consequence, there is
no difference between $\Gamma^{\rm (3loop)}[\phi,D,V_3]$ and
$\Gamma^{\rm (3loop)}[\phi,D,V_3,V_4]$ etc.~such that the 
3PI effective action captures already the 
complete answer for the self-consistent description
to this order. In contrast, at four loops the 4PI effective action 
would become relevant. To go to higher loop-order would be somewhat 
academic from the point of view of calculational feasibility
and we will concentrate on 4PI effective actions in the following. 

To present the argument we
will first consider a simple generic scalar model with cubic and
quartic interactions. The formal generalization to fermionic and
gauge fields is straightforward, and in Sec.~\ref{sec:nonabgauge} 
the construction is done
for $SU(N)$ gauge theories with fermions. 
We use here a concise notation where Latin indices represent all 
field attributes, numbering real field components and their internal
as well as space-time labels, and sum/integration over repeated 
indices is implied.
We consider the classical action 
\beq
S[\varphi] = \frac{1}{2} \varphi_i\, iD_{0,ij}^{-1}\, \varphi_j - \frac{g}{3 !}
V_{03,ijk} \varphi_i\varphi_j\varphi_k
- \frac{g^2}{4 !} V_{04,ijkl} \varphi_i\varphi_j\varphi_k\varphi_l \, ,
\label{eq:action}
\eeq
where we scaled out a constant $g$ for later convenience.
The generating functional for Green's functions in the
presence of quadratic, cubic and quartic source terms is
given by:
\bea
Z[J,R,R_3,R_4] &=& \exp\left(i W[J,R,R_3,R_4]\right) \nonumber\\
&=& \int \mathcal{D}\varphi 
\exp\Big\{ i \Big( S[\varphi] + J_i\, \varphi_i
+ \frac{1}{2} R_{ij}\, \varphi_i \varphi_j \label{eq:Z}\\
&&
+\frac{1}{3 !} R_{3,ijk}\, \varphi_i \varphi_j \varphi_k
+\frac{1}{4 !} R_{4,ijkl}\, \varphi_i \varphi_j \varphi_k \varphi_l
\Big)\Big\} \, . \nonumber
\eea
The generating functional for connected Green's functions, $W$, 
can be used to define the connected two-point ($D$), three-point ($D_3$) and
four-point function ($D_4$) in the presence of the sources,
\bea
\frac{\delta W}{\delta J_i} &=& \phi_i \, ,  \\
\frac{\delta W}{\delta R_{ij}} &=& \frac{1}{2}
\left( D_{ij} + \phi_i \phi_j \right) \, , \\
\frac{\delta W}{\delta R_{3,ijk}} &=& \frac{1}{6}
\left( D_{3,ijk} + D_{ij}\, \phi_k + D_{ki}\, \phi_j + D_{jk}\, \phi_i 
+ \phi_i\phi_j\phi_k\right) \, , \\
\frac{\delta W}{\delta R_{4,ijkl}} &=& \frac{1}{24}
( D_{4,ijkl} + [D_{3,ijk}\, \phi_l + 3\, {\rm perm.}] 
+ [D_{ij} D_{kl} + 2\, {\rm perm.}] \nonumber \\
&&
+ [D_{ij}\, \phi_k \phi_l + 5\, {\rm perm.}] 
+ \phi_i\phi_j\phi_k\phi_l ) \, .
\label{eq:Wderiv}
\eea
We denote the proper three-point and four-point vertices
by $g V_3$ and $g^2 V_4$, respectively, and define\footnote{
In terms of the standard one-particle irreducible effective action
$\Gamma[\phi]=W[J]-J \phi$ this corresponds to
\bea
g V_{3} = - \frac{\delta^3 \Gamma[\phi]}{\delta \phi\delta \phi\delta \phi} 
\quad , \quad g^2 V_{4} = 
- \frac{\delta^4 \Gamma[\phi]}{\delta \phi\delta \phi\delta \phi\delta \phi}
\nonumber .
\eea
Here it is useful to note that in terms of the connected 
Green's functions $D_n$ one has
\bea
\frac{\delta^2 W[J]}{\delta J \delta J} &=& i\, D \qquad , \qquad
\frac{\delta^2 \Gamma[\phi]}{\delta \phi \delta \phi}\,\, 
= \,\, i\, D^{-1} \, ,
\nonumber \\
\frac{\delta^3 W[J]}{\delta J \delta J \delta J} &=& -\, D_3 
\,\, =\,\, -i\, D^3\, 
\frac{\delta^3 \Gamma[\phi]}{\delta \phi \delta \phi \delta \phi}\, .
\nonumber \\
\frac{\delta^4 W[J]}{\delta J \delta J \delta J \delta J} &=& -i\, D_4 
\,\, =\,\, D^4\,  
\frac{\delta^4 \Gamma[\phi]}{\delta \phi \delta \phi \delta \phi \delta \phi}
+ 3 i\, D^5 
\left(\frac{\delta^3 \Gamma[\phi]}{\delta \phi \delta \phi \delta \phi}\right)^2\, .
\nonumber
\eea
} 
\bea
D_{3,ijk} &=& - i g\, D_{ii'}D_{jj'}D_{kk'} V_{3,i'j'k'}  \label{eq:V3def}
\, , \\
D_{4,ijkl} &=& - i g^2\, D_{ii'}D_{jj'}D_{kk'}D_{ll'} V_{4,i'j'k'l'}
\nonumber\\
&& + g^2\, (D_{ii'}D_{jj'}D_{k'u'}D_{w'l}D_{v'k} 
+ D_{ii'}D_{j'u'}D_{k'l}D_{jv'}D_{w'k} 
\nonumber\\
&& + D_{ii'}D_{j'u'}D_{k'k}D_{jv'}D_{l'l})
V_{3,i'j'k'} V_{3,u'v'w'} \, . 
\label{eq:V4def}
\eea
The effective action is obtained as the Legendre transform of
$W[J,R,R_3,R_4]$:
\bea
\Gamma[\phi,D,V_3,V_4] &=& W 
-\frac{\delta W}{\delta J_i} J_i - \frac{\delta W}{\delta R_{ij}} R_{ij}
\nonumber\\
&& - \frac{\delta W}{\delta R_{3,ijk}} R_{3,ijk} 
- \frac{\delta W}{\delta R_{4,ijkl}} R_{4,ijkl} \, .
\label{eq:LT}
\eea
For vanishing sources one observes from (\ref{eq:LT}) the stationarity 
conditions
\beq
\frac{\delta \Gamma}{\delta \phi} = \frac{\delta \Gamma}{\delta D}
= \frac{\delta \Gamma}{\delta V_3} = \frac{\delta \Gamma}{\delta V_4} = 0 \, ,
\label{eq:statcond}
\eeq
which provide the equations of motion for $\phi$, $D$, $V_3$ and $V_4$.

\subsection{$\Gamma[\phi,D,V_3,V_4]$ up to four-loop or
${\mathcal O}(g^6)$ corrections
}
\label{sec:compu}

Since the Legendre transforms employed in (\ref{eq:LT}) can be
equally performed subsequently, a most convenient 
computation of $\Gamma[\phi,D,V_3,V_4]$ 
starts from the 2PI effective action $\Gamma[\phi,D]$~\cite{4PI}.
The exact 2PI effective action can be written as~\cite{Cornwall:1974vz}:
\beq
\Gamma[\phi,D] = S[\phi] + \frac{i}{2} \Tr \ln D^{-1}
+ \frac{i}{2} \Tr\, D_{0}^{-1}(\phi) D 
+ \Gamma_{2}[\phi,D] 
+ {\rm const} \, ,
\label{eq:2PIdef}
\eeq
with the field-dependent inverse classical propagator
\beq
i D_{0}^{-1}(\phi) = \frac{\delta^2 S[\phi]}{\delta\phi \delta\phi} \, .
\eeq
To simplify the presentation, we use in the following a
symbolic notation which suppresses indices and summation or
integration symbols (suitably regularized). 
In this notation the inverse classical propagator
reads
\beq
i D_{0}^{-1}(\phi) = i D_{0}^{-1} 
- g \phi V_{03}  
-\frac{1}{2} g^2 \phi^2 V_{04}  \, , 
\eeq
and to three-loop order one has\footnote{Note that
for $\phi \not = 0$, in the phase with spontaneous symmetry breaking,
$\phi \sim \mathcal{O}(1/g)$, and the three-loop result 
(\ref{eq:2PI3loop}) takes into account the contributions up to
order $g^6$.} 
\bea
\Gamma_2 [\phi,D] &=& - \frac{1}{8} g^2 D^2 V_{04}  
+ \frac{i}{12} D^3 (g V_{03} + g^2 \phi V_{04})^2 
+ \frac{i}{48} g^4 D^4 V_{04}^2  
\nonumber\\
&& + \frac{1}{8} g^2 D^5 (g V_{03} + g^2 \phi V_{04})^2 V_{04} 
- \frac{i}{24} D^6 (g V_{03} + g^2 \phi V_{04})^4 
\nonumber\\[0.13cm]
&& +\, \mathcal{O}\left( 
g^n (g^2 \phi)^m |_{n+m=6}
\right)   \, ,
\label{eq:2PI3loop}
\eea
for $n,m=0,\ldots,6$. We emphasize that the exact $\phi$-dependence of 
$\Gamma_2[\phi,D]$ can be written as a function of the combination 
$(g V_{03} + g^2 \phi V_{04})$.
In order to obtain the vertex 2PI effective action $\Gamma[\phi,D,V_3,V_4]$
from $\Gamma [\phi,D]$, one can exploit 
that the cubic and quartic source terms 
$\sim R_3$ and $\sim R_4$ appearing in (\ref{eq:Z}) can be conveniently 
combined with the vertices $g V_{03}$ and $g^2 V_{04}$
by the replacement:
\beq
g V_{03} \,\,\to\,\, g V_{03} - R_3 \equiv g \tilde{V}_3 
\quad , \quad g^2 V_{04} \,\,\to\,\, g^2 V_{04} - R_4 
\equiv g^2 \tilde{V}_4 \,\, .
\eeq
The 2PI effective action with the modified interaction is given by
\beq
\Gamma_{\tilde{V}}[\phi,D] = W[J,R,R_3,R_4] 
-\frac{\delta W}{\delta J} J - \frac{\delta W}{\delta R} R \, .
\eeq
Since
\beq
\frac{\delta \Gamma_{\tilde{V}}}{\delta R_3} = 
\frac{\delta W}{\delta R_3} \qquad ,\qquad
\frac{\delta \Gamma_{\tilde{V}}}{\delta R_4} = 
\frac{\delta W}{\delta R_4} \,\, ,
\eeq
one can express the remaining Legendre transforms, leading to 
$\Gamma[\phi,D,V_3,V_4]$, in terms of the vertices $\tilde{V}_{3}$,
$\tilde{V}_{4}$ and
$V_{03}$, $V_{04}$:   
\bea
\lefteqn{\Gamma[\phi,D,V_3,V_4] = \Gamma_{\tilde{V}}[\phi,D] 
- \frac{\delta \Gamma_{\tilde{V}}[\phi,D]}{\delta R_{3}} R_{3}
- \frac{\delta \Gamma_{\tilde{V}}[\phi,D]}{\delta R_{4}} R_{4}}
\nonumber\\
&=& \Gamma_{\tilde{V}}[\phi,D] 
- \frac{\delta \Gamma_{\tilde{V}}[\phi,D]}{\delta \tilde{V}_{3}} 
(\tilde{V}_3 - V_{03})
- \frac{\delta \Gamma_{\tilde{V}}[\phi,D]}{\delta \tilde{V}_{4}} 
(\tilde{V}_4 - V_{04})  \, .
\label{eq:VGcomp}
\eea
What remains to be done is expressing 
$\tilde{V}_3$ and $\tilde{V}_4$ in terms of $V_3$ and $V_4$.
On the one hand, from (\ref{eq:Wderiv}) and the 
definitions (\ref{eq:V3def}) and (\ref{eq:V4def}) one has
\bea
\frac{\delta \Gamma_{\tilde{V}}[\phi,D]}{g \delta \tilde{V}_{3}}
&=& - \frac{1}{6}
\left( - i g\, D^3 V_3 + 3 D \phi  
+ \phi^3 \right)\, ,
\label{eq:DGV3}\\
\frac{\delta \Gamma_{\tilde{V}}[\phi,D]}{g^2 \delta \tilde{V}_{4}}
&=& - \frac{1}{24}
\left( - i g^2\, D^4 V_4
- 3 g^2 D^5 V_3^2
- 4 i g\, D^3 V_3 \phi 
+ 3 D^2 \right. 
\nonumber\\
&& \left. + 6 D \phi^2  
+ \phi^4 \right) \, . 
\label{eq:DGV4}
\eea
On the other hand, from the expansion of the 2PI effective action
to three-loop order with (\ref{eq:2PI3loop}) one 
finds\footnote{Note that since the exact $\phi$-dependence of 
$\Gamma_2[\phi,D]$ can be written as a function of 
$(g V_{03} + g^2 \phi V_{04})$, the parametrical dependence of
the higher order terms in 
the variation of (\ref{eq:2PI3loop}) 
with respect to $(g V_{03})$ is given by 
$\mathcal{O}(g^{n} (g^2 \phi)^m |_{n+m=5})$ (cf.~(\ref{eq:2PIV3})). 
}  
\bea
\frac{\delta \Gamma_{\tilde{V}}[\phi,D]}{g \delta \tilde{V}_{3}}
&=& - \frac{1}{6} \phi^3 - \frac{1}{2} D \phi + \frac{i}{6} D^3 
(g \tilde{V}_3 + g^2 \phi \tilde{V}_4) 
\nonumber \\
&& + \frac{1}{4} g^2 D^5 (g \tilde{V}_3 + g^2 \phi \tilde{V}_4) \tilde{V}_4 
- \frac{i}{6} D^6 (g \tilde{V}_3 + g^2 \phi \tilde{V}_4)^3 
\nonumber\\[0.13cm]
&& +\, \mathcal{O}\left( 
g^{n} (g^2 \phi)^m |_{n+m=5}
\right) \, ,
\label{eq:2PIV3}\\[0.13cm]
\frac{\delta \Gamma_{\tilde{V}}[\phi,D]}{g^2 \delta \tilde{V}_{4}}
&=& -\frac{1}{24} \phi^4 - \frac{1}{4} D \phi^2 - \frac{1}{8} D^2 
+ \frac{i}{6} D^3 \phi (g \tilde{V}_3 + g^2 \phi \tilde{V}_4)
\nonumber\\
&& + \frac{i}{24} g^2 D^4 \tilde{V}_4 
+ \frac{1}{4} g^2 D^5 \phi 
(g \tilde{V}_3 + g^2 \phi \tilde{V}_4) \tilde{V}_4
\nonumber\\
&& 
+ \frac{1}{8} D^5 (g \tilde{V}_3 + g^2 \phi \tilde{V}_4)^2 
- \frac{i}{6} D^6 \phi (g \tilde{V}_3 + g^2 \phi \tilde{V}_4)^3
\nonumber\\[0.13cm]
&& +\, \mathcal{O}\left( 
g^{n-2} (g^2 \phi)^m |_{n+m=6}
\right) \, .
\label{eq:2PIV4}
\eea
Comparing (\ref{eq:2PIV3}) and (\ref{eq:DGV3}) yields
\bea
g V_3 &=& (g \tilde{V}_3 + g^2 \phi \tilde{V}_4) 
- \frac{3}{2} i g^2 D^2 (g \tilde{V}_3 + g^2 \phi \tilde{V}_4) \tilde{V}_4
- D^3 (g \tilde{V}_3 + g^2 \phi \tilde{V}_4)^3 
\nonumber\\
&& +\, \mathcal{O}\left( 
g^{n} (g^2 \phi)^m |_{n+m=5}
\right) \, .
\label{eq:V3pre}
\eea
Similarly, for $V_4$ comparing (\ref{eq:2PIV4}) and (\ref{eq:DGV4}), 
and using (\ref{eq:V3pre}) one finds 
\beq
g^2 V_4 = g^2 \tilde{V}_4  + \mathcal{O}\left( 
g^{n-2} (g^2 \phi)^m |_{n+m=6}
\right) \, .
\label{eq:V3res} 
\eeq
This can be used to invert the above relations as 
\bea
g \tilde{V}_3 + g^2 \phi \tilde{V}_4 &=& g V_3 + \frac{3}{2} i g^3 D^2 V_3 V_4
+ g^3 D^3 V_3^3 
+ \mathcal{O}\left(g^5 \right) \, ,  \\[0.14cm]
g^2 \tilde{V}_4 &=& g^2 V_4   + \mathcal{O}\left( g^4 \right) \, .
\eea
Plugging this into (\ref{eq:VGcomp}) and expressing the free, the one-loop and
the $\Gamma_2$ parts in terms of $V_{3}$ and $V_{4}$ as well as
$V_{03}$ and $V_{04}$, one obtains from a straightforward calculation:
\beq
\Gamma[\phi,D,V_3,V_4] = S[\phi] + \frac{i}{2} \Tr \ln D^{-1}
+ \frac{i}{2} \Tr\, D_0^{-1}(\phi) D + \Gamma_2[\phi,D,V_3,V_4] \, ,
\label{eq:V2PIeffact}
\eeq
with 
\bea
\Gamma_2[\phi,D,V_3,V_4] &=& \Gamma_2^0[\phi,D,V_3,V_4]
+ \Gamma_2^{\rm int}[D,V_3,V_4] \, , \\[0.2cm]
\Gamma_2^0[\phi,D,V_3,V_4] &=& - \frac{1}{8} g^2 D^2 V_{04}
+ \frac{i}{6} g D^3 V_3 (g V_{03} + g^2 \phi V_{04})
\nonumber\\
&&
+ \frac{i}{24} g^4 D^4 V_4 V_{04}
+ \frac{1}{8} g^4 D^5 V_3^2 V_{04} \, ,
\label{eq:G20phi} \\
\Gamma_2^{\rm int}[D,V_3,V_4] &=& - \frac{i}{12} g^2 D^3 V_3^2 
- \frac{i}{48} g^4 D^4 V_4^2  
- \frac{i}{24} g^4 D^6 V_3^4 
+ \mathcal{O}\left(g^6\right) . \qquad
\label{eq:G2int}
\eea
The diagrammatic representation of these results is given
in Figs.~\ref{fig:ym3loop1} and~\ref{fig:ym3loop3} of 
Sec.~\ref{sec:calcSUN}. There the equivalent calculation is
done for a $SU(N)$ gauge theory and one has to replace
the propagator lines and vertices of the figures by the corresponding scalar 
propagator and vertices. Note that for the scalar theory
the thick circles represent the dressed three-vertex $g V_3$ 
and four-vertex $g^2 V_4$, respectively, while the small circles denote
the corresponding {\em effective} classical three-vertex 
$g V_{03} + g^2 \phi V_{04}$ and classical four-vertex
$g^2 V_{04}$. As a consequence, the diagrams
look the same in the absence of spontaneous symmetry breaking,
indicated by a vanishing field expectation value $\phi$. 

In (\ref{eq:V2PIeffact}), the action $S[\phi]$
and $D_0$ depend on the classical vertices as before.
The expression for $\Gamma_2^0$, which includes all terms of 
$\Gamma_2$ that depend on the classical vertices, is valid to
all orders: $\Gamma_2^{\rm int}$ contains no explicit dependence
on the field $\phi$ or the classical vertices $V_{03}$ and $V_{04}$,
independent of the approximation for the 4PI effective action. This 
can be straightforwardly observed from (\ref{eq:VGcomp}), where the complete
(linear) dependence of $\Gamma$ on $V_{03}$ and $V_{04}$ is explicit,
together with (\ref{eq:DGV3}) and (\ref{eq:DGV4}).

\subsection{Self-consistently complete loop/coupling expansion}
\label{sec:selfcomplete}

As pointed out in Sec.~\ref{sec:npieffectiveactions},
for applications it is often desirable to obtain a self-consistently 
complete description, which to a given order of a loop or coupling
expansion determines the $n$PI effective action 
$\Gamma[\phi,D,V_3,V_4,\ldots,V_n]$ for arbitrarily 
high $n$. Despite the complexity of a general $n$PI effective action
such a description can be obtained in practice because 
of the equivalence hierarchy displayed in Eq.~(\ref{eq:hierarchy}):
Typically the 2PI, 3PI or maybe the 4PI effective action  
captures already the complete answer for the self-consistent 
description to the desired/computationally feasible order of 
approximation~\cite{Cornwall:1974vz,4PI,Calzetta:1988cq}. 
Higher effective actions, which are relevant 
beyond four-loop order, may not be entirely irrelevant in the 
presence of sources describing complicated initial conditions 
for nonequilibrium evolutions. However, their discussion would 
be rather academic from the point of view of calculational
feasibility and we will concentrate
on up to four-loop corrections or ${\mathcal O}(g^6)$
in the following. Below we will not explicitly write in addition to the
loop-order the order of the coupling $g$, which is straightforward
as detailed above in Sec.~\ref{sec:compu}.

To show (\ref{eq:hierarchy}) we will first observe that
to one-loop order all $n$PI effective actions agree in the
absence of sources. The standard one-loop expression for the 1PI 
effective action reads~\cite{IZ},
\beq
\Gamma^{\rm (1loop)}[\phi] = S[\phi] + \frac{i}{2} \Tr \ln D_0^{-1}(\phi) \, . 
\label{eq:1pioneloop}
\eeq
For the 2PI effective action one finds from (\ref{eq:2PIdef}) and 
(\ref{eq:2PI3loop}) up to an irrelevant constant
\bea
\Gamma^{\rm (1loop)}[\phi,D] = S[\phi] + \frac{i}{2} \Tr \ln D^{-1}
+ \frac{i}{2} \Tr\, D_0^{-1}(\phi) D \, .
\label{eq:2pioneloop}
\eea  
The absence of sources 
(since $\delta \Gamma^{\rm (1loop)}[\phi,D]/\delta D =  - R_2/2$,
cf.~Sec.~\ref{sec:highereffective actions}~\cite{Cornwall:1974vz}) 
corresponds to $D$ given by
\beq
\frac{\delta \Gamma^{\rm (1loop)}[\phi,D]}{\delta D} = 0
\qquad \Rightarrow \qquad 
D^{-1} = D_0^{-1}(\phi) \, .
\eeq
Using this result in Eq.~(\ref{eq:2pioneloop}) and 
comparing\footnote{Up to irrelevant constants, 
which are given by the choice of normalization for $\Gamma$.
($\Tr D_0^{-1} D_0 = \Tr {\bf 1} = \mbox{const}$.)} 
with (\ref{eq:1pioneloop}) one has
\beq
\Gamma^{\rm (1loop)}[\phi,D] = \Gamma^{\rm (1loop)}[\phi] \, ,
\eeq 
in the absence of sources. The equivalence with the
one-loop 3PI and 4PI effective actions can be explicitly observed from 
the results of Sec.~\ref{sec:compu}. 
In order to obtain the 3PI expressions we could 
directly set the source $R_4\equiv 0$ from the beginning in 
the computation of that section such that there is no
dependence on $V_4$. Equivalently, we can note from 
Eqs.~(\ref{eq:V2PIeffact})--(\ref{eq:G2int}) that already the 4PI 
effective action to this order simply agrees with (\ref{eq:2pioneloop}).
As a consequence, it carries no dependence on $V_3$ and $V_4$, i.e.
\beq
\Gamma^{\rm (1loop)}[\phi,D,V_3,V_4] 
= \Gamma^{\rm (1loop)}[\phi,D,V_3] 
= \Gamma^{\rm (1loop)}[\phi,D] \, .
\eeq  
For the one-loop case it remains to be shown that in addition
\beq
\Gamma^{\rm (1loop)}[\phi,D,V_3,V_4,\ldots,V_n]
= \Gamma^{\rm (1loop)}[\phi,D,V_3,V_4]
\label{eq:equiv4vn} 
\eeq
for arbitrary $n \ge 5$. For this we note that the 
number $I$ of internal lines in a given loop diagram is given
by the number $v_3$ of proper 3-vertices, the number $v_4$ of 
proper 4-vertices, \ldots, the number $v_n$ of 
proper n-vertices in terms of the standard relation: 
\beq
2 I = 3 v_3 + 4 v_4 + 5 v_5 \ldots + n v_n \, ,
\eeq
where $v_3 + v_5 + v_7 + \ldots$ has to be even.
Similarly, the number $L$ of loops in such a diagram is
\bea
L &=& I - v_3 - v_4 - v_5 \ldots - v_n + 1 \nonumber\\
&=& \frac{1}{2} v_3 + v_4 + \frac{3}{2} v_5 \ldots
+ \frac{n - 2}{2} v_n + 1  \,\, .
\label{eq:numberofloops}
\eea
The equivalence~(\ref{eq:equiv4vn}) follows from the fact
that for~$L=1$ Eq.~(\ref{eq:numberofloops}) implies that 
$\Gamma^{\rm (1loop)}[\phi,D,V_3,V_4,\ldots,V_n]$ 
cannot depend in particular on $V_5, \ldots V_n$.\footnote{Note
that we consider here theories where there is no 
classical 5-vertex or higher, whose presence would lead to a trivial
dependence for the classical action and propagator.}

The two-loop equivalence of the 2PI and higher effective
actions follows along the same lines. According to
(\ref{eq:V2PIeffact})--(\ref{eq:G2int}) the 4PI effective
action to two-loop order is given by:
\bea
\Gamma^{\rm (2loop)}[\phi,D,V_3,V_4] 
&\!=\!& S[\phi] + \frac{i}{2} \Tr \ln D^{-1}
+ \frac{i}{2} \Tr\, D_0^{-1}(\phi) D \nonumber\\
&& +\, \Gamma_2^{\rm (2loop)}[\phi,D,V_3,V_4] 
\, , \label{eq:4pitwoloop} \\
\Gamma_2^{\rm (2loop)}[\phi,D,V_3,V_4] &\!=\!&
- \frac{1}{8} g^2 D^2 V_{04}
+ \frac{i}{6} g D^3 V_3 (g V_{03} + g^2 \phi V_{04})
- \frac{i}{12} g^2 D^3 V_3^2 \nonumber \, .
\eea
There is no dependence on $V_4$ to this order and, following 
the discussion above, there is no dependence on
$V_5,\ldots,V_n$ according to (\ref{eq:numberofloops}) for
$L=2$. Consequently, 
\beq
\Gamma^{\rm (2loop)}[\phi,D,V_3,V_4,\ldots,V_n] =
\Gamma^{\rm (2loop)}[\phi,D,V_3,V_4] =
\Gamma^{\rm (2loop)}[\phi,D,V_3] \, ,
\eeq
for arbitrary $n$ in the absence of sources. The latter yields
\beq
\frac{\delta \Gamma^{\rm (2loop)}[\phi,D,V_3]}{\delta V_3} = 
\frac{\delta \Gamma^{\rm (2loop)}_2[\phi,D,V_3]}{\delta V_3} = 0
\quad \Rightarrow \quad 
g V_3 = g V_{03} + g^2 \phi V_{04} \, ,
\eeq
which can be used in (\ref{eq:4pitwoloop}) to show in addition
the equivalence of the
3PI and 2PI effective actions (cf.~Eq.~(\ref{eq:2PI3loop})) to 
this order:
\bea
\Gamma_2^{\rm (2loop)} [\phi,D,V_3] &=& - \frac{1}{8} g^2 D^2 V_{04}  
+ \frac{i}{12} D^3 (g V_{03} + g^2 \phi V_{04})^2 \nonumber\\
&=& \Gamma_2^{\rm (2loop)} [\phi,D] \, ,
\label{eq:2pitwoloop}
\eea 
for vanishing sources. The {\em inequivalence} of the 2PI with 
the 1PI effective action to this order,
\beq
\Gamma^{\rm (2loop)} [\phi,D] \not = \Gamma^{\rm (2loop)}[\phi] \, ,
\eeq
follows from using the result of 
$\delta\Gamma_2^{\rm (2loop)} [\phi,D]/\delta D = 0$ for $D$
in (\ref{eq:2pitwoloop}) in a straightforward way\footnote{
Here $\Gamma^{\rm (2loop)} [\phi,D]$ includes e.g.~the 
summation of an infinite series of so-called ``bubble''
diagrams, which form the basis of mean-field or Hartree-type
approximations, and clearly go beyond a perturbative two-loop
approximation $\Gamma^{\rm (2loop)}[\phi]$.}~\cite{Cornwall:1974vz}.

In order to show the three-loop equivalence of the 3PI and
higher effective actions, we first note from 
(\ref{eq:V2PIeffact})--(\ref{eq:G2int}) that the 4PI
effective action to this order yields $V_4 = V_{04}$ in
the absence of sources:
\beq
\frac{\delta \Gamma^{\rm (3loop)}[\phi,D,V_3,V_4]}{\delta V_4} = 
\frac{\delta \Gamma^{\rm (3loop)}_2[\phi,D,V_3,V_4]}{\delta V_4} = 
\frac{i}{24} g^4 D^4 \left( V_{04} - V_4 \right) = 0 \, .
\eeq
Constructing the 3PI effective action to three-loop would mean to 
do the same calculation as in Sec.~\ref{sec:compu} but with
$V_4 \to V_{04}$ from the beginning ($R_4 \equiv 0$). 
The result of a classical four-vertex for the 4PI effective action
to this order, therefore, directly implies:
\beq
\Gamma^{\rm (3loop)}[\phi,D,V_3,V_4] = \Gamma^{\rm (3loop)}[\phi,D,V_3] \, ,
\label{eq:parteq} 
\eeq
for vanishing sources. To see the equivalence with a 5PI effective
action $\Gamma^{\rm (3loop)}[\phi,D,V_3,V_4,V_5]$,
we note that to three-loop order the only possible diagram
including a five-vertex requires $v_3 = v_5 = 1$ for $L=3$
in Eq.~(\ref{eq:numberofloops}). As a consequence, to this
order the five-vertex corresponds to the classical one, which 
is identically zero for the theories considered here, 
i.e.~$V_5 = V_{05} \equiv 0$. In order to obtain
that (to this order trivial) result along the lines of 
Sec.~\ref{sec:compu}, one can formally include a classical
five-vertex $V_{05}$ and observe that the three-loop
2PI effective action
admits a term $\sim D^4 V_{05} V_3$. After performing the
additional Legendre transform the result then follows
from setting $V_{05} \to 0$ in the end. 
The equivalence with $n$PI effective actions for
$n \ge 6$ can again be observed from the fact
that for~$L=3$ Eq.~(\ref{eq:numberofloops}) implies 
no dependence on $V_6, \ldots V_n$. In addition to
(\ref{eq:parteq}), we therefore have for
arbitrary $n \ge 5$:
\beq
\Gamma^{\rm (3loop)}[\phi,D,V_3,V_4,\ldots,V_n] =
\Gamma^{\rm (3loop)}[\phi,D,V_3,V_4] \, .
\eeq
The {\em inequivalence} of the three-loop 3PI and 
2PI effective actions can be readily observed
from (\ref{eq:V2PIeffact})--(\ref{eq:G2int}) and
(\ref{eq:parteq}):
\bea
\frac{\delta \Gamma^{\rm (3loop)}[\phi,D,V_3]}{\delta V_3} = 0
\quad \Rightarrow \quad 
g V_3 = g \left(V_{03} + g \phi V_{04}\right) - g^3 D^3 V_3^3 \, .
\eea 
Written iteratively, the above self-consistent equation for $V_3$
sums an infinite number of contributions in terms of the
classical vertices. As a consequence, the three-loop 3PI result 
can be written as an infinite series of diagrams for the
corresponding 2PI effective action, which
clearly goes beyond $\Gamma^{\rm (3loop)}[\phi,D]$ 
(cf.~Eq.~(\ref{eq:2PI3loop})):
\beq
\Gamma^{\rm (3loop)}[\phi,D,V_3] \not = \Gamma^{\rm (3loop)}[\phi,D] \, .
\eeq
The importance of such an infinite
summation will be discussed for the case of gauge theories below.

\section{Nonabelian gauge theory with fermions} 
\label{sec:nonabgauge}

We consider a $SU(N)$ gauge theory with $N_f$ flavors of
Dirac fermions with classical action
\bea
\lefteqn{
S_{\rm eff} = S + S_{\rm gf} + S_{\rm FPG} }  \\
&=& \int d^4x \Big( 
- \frac{1}{4} F_{\mu\nu}^a F^{\mu\nu\, a}
- \frac{1}{2 \xi} \left({\cal G}^a[A] \right)^2
- \bar{\psi} (-i D\slash) \psi
- \bar{\eta}^a \partial_\mu \left(D^\mu \eta \right)^a
\Big)\,\, , \nonumber
\eea
where $\psi$ ($\bar{\psi}$), $A$ and $\eta$ ($\bar{\eta}$) denote
the (anti-)fermions, gauge and 
\mbox{(anti-)ghost} fields, respectively. 
The color indices in the adjoint representation are 
$a,b,\ldots = 1,\ldots,N^2-1$, while those for the fundamental representation
will be denoted by $i,j,\ldots$ and run from $1$ to $N$. The gauge-fixing term
${\cal G}^a[A]$ is ${\cal G}^a = 
\partial^\mu A_\mu^a$ for covariant gauges. Here
\bea
F_{\mu\nu}^a &=& \partial_\mu A_\nu^a - \partial_\nu A_\mu^a
- g f^{abc} A_\mu^b A_\nu^c \, ,  \\ 
\left(D^\mu \eta \right)^a &=& \partial^\mu \eta^a - g f^{abc} A^{\mu\, b} 
\eta^c
\, ,\\
D\slash &=& \gamma^\mu \left(\partial_\mu + i g A^a_\mu t^a \right) \, , 
\eea
where $[t^a,t^b] = i f^{abc} t^c$, $\tr(t^a t^b) = \delta^{ab}/2$. For QCD,
$t^a = \lambda^a/2$ with the Gell-Mann matrices $\lambda^a$
($a=1,\ldots,8$). We will suppress Dirac and flavor indices in
the following. It is convenient to write $S_{\rm eff}$ in the compact form: 
\bea
S_{\rm eff} &=& \frac{1}{2} \int_{x y} A^{\mu\, a}(x)\,
i D^{-1\, ab}_{0\, \mu\nu}(x,y) A^{\nu\, b}(y) 
+ \int_{x y} \bar{\eta}^a(x)\, i G_0^{-1\, ab}(x,y) \eta^b(y)
\nonumber\\
&+& \int_{x y} \bar{\psi}_i(x)\, i \Delta_{0\, ij}^{-1}(x,y) \psi_j(y)
\nonumber\\
&-& \frac{1}{6}\, g \int_{x y z} V_{03\, \mu\nu\gamma}^{abc}(x,y,z)
A^{\mu\, a}(x) A^{\nu\, b}(y) A^{\gamma\, c}(z) 
\nonumber\\
&-& \frac{1}{24}\, g^2 \int_{x y z w} 
V_{04\, \mu\nu\gamma\delta}^{abcd}(x,y,z,w) 
A^{\mu\, a}(x) A^{\nu\, b}(y) A^{\gamma\, c}(z) A^{\delta\, d}(w) 
\nonumber\\
&-& g \int_{x y z} V_{03\, \mu}^{{\rm (gh)}ab,c}(x,y;z)
\bar{\eta}^a(x)  \eta^b(y) A^{\mu\, c}(z)
\nonumber\\
&-& g \int_{x y z} V_{03\, \mu\, i j}^{{\rm (f)}a}(x,y;z)
\bar{\psi}_i(x)  \psi_j(y) A^{\mu\, a}(z) \, ,
\label{eq:Seff}
\eea
with the free inverse fermion, ghost and gluon propagator
in covariant gauges given by
\bea
i \Delta^{-1}_{0\, ij}(x,y) &=& i \partial_x \!\slash 
\delta_{ij} \delta_\C (x-y) \, ,
\\[0.14cm]
i G^{-1\, ab}_{0}(x,y) &=& - \square_x \delta^{ab} \delta_\C (x-y) \, ,
\\[0.14cm]
i D^{-1\, ab}_{0\, \mu\nu}(x,y) 
&=& \left[g_{\mu\nu}\, \square - \left( 1 - \xi^{-1}\right) 
\partial_\mu \partial_\nu \right]_x \delta^{ab} \delta_\C (x-y) 
 \, ,
\eea
where we have taken the fermions to be massless.
The tree-level vertices read in coordinate space: 
\bea
V_{03\, \mu\nu\gamma}^{abc}(x,y,z) &=& f^{abc} \Big( 
\nonumber\\
&& g_{\mu\nu} [\delta_{\C}(y-z)\, \partial^x_\gamma\delta_{\C}(x-y)
- \delta_{\C}(x-z)\, \partial^y_\gamma\delta_{\C}(y-x)]
\nonumber\\[0.13cm]
&+& 
 g_{\mu\gamma} [\delta_{\C}(x-y)\, \partial^z_\nu\delta_{\C}(z-x)
- \delta_{\C}(y-z)\, \partial^x_\nu\delta_{\C}(x-z)]
\nonumber\\
&+& 
 g_{\nu\gamma} [\delta_{\C}(x-z)\, \partial^y_\mu\delta_{\C}(y-x)
- \delta_{\C}(x-y)\, \partial^z_\mu\delta_{\C}(z-x)]
\Big) \, ,\nonumber\\
\label{eq:vert1}\\ 
V_{04\, \mu\nu\gamma\delta}^{abcd}(x,y,z,w) &=& \Big(
f^{abe}f^{cde} [ g_{\mu\gamma}g_{\nu\delta} - g_{\mu\delta}g_{\nu\gamma}]
\nonumber\\
&+& f^{ace}f^{bde} [ g_{\mu\nu}g_{\gamma\delta} - g_{\mu\delta}g_{\nu\gamma}]
+ f^{ade}f^{cbe} [ g_{\mu\gamma}g_{\delta\nu} - g_{\mu\nu}g_{\gamma\delta}]  
\Big)
\nonumber\\
&& \delta_{\C}(x-y)\delta_{\C}(x-z)\delta_{\C}(x-w) \, , 
\\[0.14cm]
V_{03\, \mu}^{{\rm (gh)}ab,c}(x,y;z) &=& - f^{abc} \partial_\mu^x 
\delta_{\C}(x-z)\delta_{\C}(y-z) \, ,
\\[0.14cm]
V_{03\, \mu\, i j}^{{\rm (f)}a}(x,y;z) &=& \gamma_\mu t^a_{ij}
\delta_{\C}(x-z)\delta_{\C}(z-y) \, .
\label{eq:fermvertex}
\eea
Note that $V_{03, abc}^{\mu\nu\gamma}(x,y,z)$
is symmetric under exchange of $(\mu,a,x) \leftrightarrow (\nu,b,y) 
\leftrightarrow (\gamma,c,z)$. Likewise, 
$V_{04, abcd}^{\mu\nu\gamma\delta}(x,y,z,w)$ is symmetric in its space-time
arguments and under exchange of $(\mu,a) \leftrightarrow (\nu,b) 
\leftrightarrow (\gamma,c) \leftrightarrow (\delta,d)$.   

\subsection{Source terms}

In addition to the linear and bilinear source terms, which
are required for a construction of the 2PI effective action,  
following Sec.~\ref{sec:highereffective actions} 
we add cubic and quartic source terms to
(\ref{eq:Seff}): 
\bea
S'_{\rm source} &=& 
 \frac{1}{6} \int_{x y z} R_{3\, \mu\nu\gamma}^{abc}(x,y,z)
A^{\mu\, a}(x) A^{\nu\, b}(y) A^{\gamma\, c}(z) 
\nonumber\\
&+& \frac{1}{24} \int_{x y z w} 
R_{4\, \mu\nu\gamma\delta}^{abcd}(x,y,z,w) 
A^{\mu\, a}(x) A^{\nu\, b}(y) A^{\gamma\, c}(z) A^{\delta\, d}(w) 
\nonumber\\
&+& \int_{x y z} R_{3\, \mu}^{{\rm (gh)}ab,c}(x,y;z)
\bar{\eta}^a(x)  \eta^b(y) A^{\mu\, c}(z)
\nonumber\\
&+& \int_{x y z} R_{3\, \mu\, i j}^{{\rm (f)}a}(x,y;z)
\bar{\psi}_i(x)  \psi_j(y) A^{\mu\, a}(z)  \, ,
\label{eq:sources}
\eea
where the sources $R_{3,4}$ obey the same symmetry properties
as the corresponding classical vertices $V_{03}$ and $V_{04}$ 
discussed above.
The definition of the corresponding three- and four-vertices 
follows Sec.~\ref{sec:highereffective actions}. In particular, we have
for the vertices involving Grassmann fields:
\bea
\frac{\delta W}{\delta R_{3\, \mu}^{{\rm (gh)}ab,c}(x,y;z)}
&=& - i g \int_{x' y' z'} D^{\mu\mu'\, c c'}(z,z') G^{b a'}(y,x') 
 \nonumber\\
&&  V_{3\, \mu'}^{{\rm (gh)}a' b' c'}(x',y';z') G^{b' a}(y',x) , 
\nonumber\\[0.2cm]
\frac{\delta W}{\delta R_{3\, \mu\, i j}^{{\rm (f)}a}(x,y;z)}
&=& - i g \int_{x' y' z'} D^{\mu\mu'\, a a'}(z,z') \Delta_{ji'}(y,x') 
 \nonumber\\ 
&& V_{3\, \mu'\, i' j'}^{{\rm (f)}a'}(x',y';z') \Delta_{j'i}(y',x) \, , 
\eea
for vanishing
`background' fields 
$\langle A \rangle = \langle \psi \rangle = \langle \bar{\psi} \rangle
= \langle \eta \rangle = \langle \bar{\eta} \rangle = 0$.

\subsection{Effective action up to four-loop or
${\mathcal O}(g^6)$ corrections}
\label{sec:calcSUN}

\begin{figure}[t]
\begin{center}
\epsfig{file=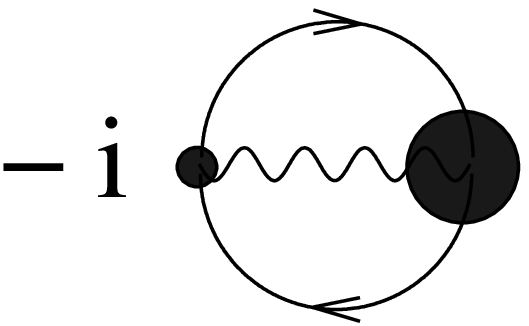,width=2.3cm}
\end{center}
\vspace*{-0.5cm}
\caption{\small Ghost/fermion part of $\Gamma_2^0$.}
\label{fig:ym3loop2}
\end{figure}

\begin{figure}[t]
\begin{center}
\epsfig{file=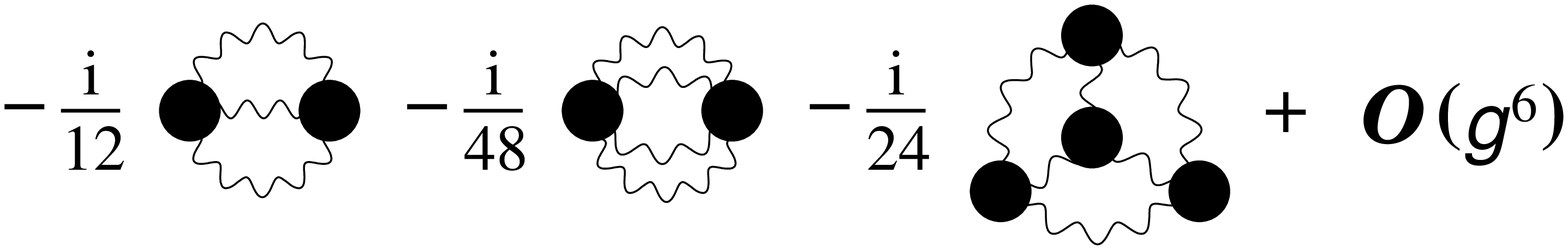,width=11.5cm}
\end{center}
\vspace*{-0.5cm}
\caption{\small The figure shows together with Fig.~\ref{fig:ym3loop4}
the diagrammatic representation of 
$\Gamma_2^{\rm int}[D,G,V_{3},V_3^{\rm (gh)},V_{4}]$
to three-loop order as given in Eq.~(\ref{eq:gamma2}). 
For the gauge theory with fermions, to this order there is in 
addition the same contribution as in
Fig.~\ref{fig:ym3loop4} with the unwiggled propagator lines
representing the fermion propagator $\Delta$ and the ghost vertex 
replaced by the corresponding fermion vertex $V_{3}^{\rm (f)}$. 
This functional contains no explicit
dependence on the classical vertices independent of the
order of approximation.}
\label{fig:ym3loop3}
\end{figure}

\begin{figure}[t]
\begin{center}
\epsfig{file=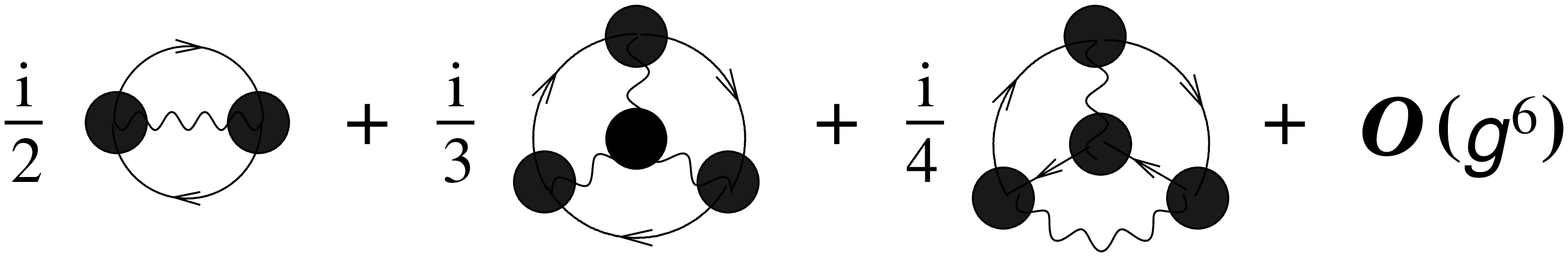,width=11.3cm}
\end{center}
\vspace*{-0.5cm}
\caption{\small Ghost/fermion part of 
$\Gamma_2^{\rm int}$ to three-loop order.}
\label{fig:ym3loop4}
\end{figure}

Consider first the standard 2PI effective action with vanishing
`background' fields,
which can be written as~\cite{Cornwall:1974vz}:
\bea
\Gamma[D,\Delta,G] 
&=& \frac{i}{2} \Tr \ln D^{-1}
+ \frac{i}{2} \Tr\, D_0^{-1} D 
- i \Tr \ln \Delta^{-1}
- i \Tr\, \Delta_0^{-1} \Delta 
\nonumber\\
&& 
- i \Tr \ln G^{-1}
- i \Tr\, G_0^{-1} G 
+ \Gamma_2[D,\Delta,G] \, 
\label{eq:exact2PI}.
\eea
Here the trace $\Tr$ includes an integration over the
time path $\C$, as well as integration over spatial 
coordinates and summation over flavor, color and Dirac indices. 
The exact expression for $\Gamma_2$ contains all 2PI diagrams
with vertices described by (\ref{eq:vert1})--(\ref{eq:fermvertex}) 
and propagator lines associated to the full connected two-point functions
$D$, $G$ and $\Delta$. In order to clear up the presentation, we will
give all diagrams including gauge and ghost propagators only.
The fermion diagrams can simply be obtained from the
corresponding ghost ones, since they have the same signs and 
prefactors\footnote{Note that to three-loop order there 
are no graphs with more than one closed ghost/fermion loop,
such that ghosts and fermions cannot appear in the same diagram
simultaneously.}. For the 2PI effective action of the gluon-ghost 
system, $\Gamma[D,G]$, to three-loop order the 2PI effective
action is given by (using the same compact notation as introduced in 
Sec.~\ref{sec:compu}):
\bea
\Gamma_2[D,G] &=& 
- \frac{1}{8} g^2 D^2 V_{04}
+ \frac{i}{12} g^2 D^3 V_{03}^2 
- \frac{i}{2} g^2 D G^2 V_{03}^{\rm (gh)\, 2}
+ \frac{i}{48} g^4 D^4 V_{04}^2
\nonumber\\
&& + \frac{1}{8} g^4 D^5 V_{03}^2 V_{04}  
- \frac{i}{24} g^4 D^6 V_{03}^4 
+ \frac{i}{3} g^4 D^3 G^3 V_{03}^{\rm (gh)\, 3}V_{03} 
\nonumber\\
&&+ \frac{i}{4} g^4 D^2 G^4 V_{03}^{\rm (gh)\, 4} 
+ \mathcal{O}\left(g^6 \right)
\, .
\label{eq:3loop2PISUN}
\eea
The result can be compared with (\ref{eq:2PI3loop}) and taking
into account an additional factor of $(-1)$ for each closed
loop involving Grassmann fields~\cite{Cornwall:1974vz}.
Here we have suppressed in the notation the dependence of $\Gamma_2[D,G]$
on the higher sources (\ref{eq:sources}). The desired effective action
is obtained by performing the remaining Legendre
transforms:
\bea
\Gamma[D,G,V_{3},V_3^{\rm (gh)},V_{4}] = \Gamma[D,G]
- \frac{\delta W}{\delta R_3} R_3 
- \frac{\delta W}{\delta R_3^{\rm (gh)}} R_3^{\rm (gh)}
- \frac{\delta W}{\delta R_4} R_4  \, .
\eea
The calculation follows the same steps as detailed in Sec.~\ref{sec:compu}.  
\begin{figure}[t]
\begin{center}
\epsfig{file=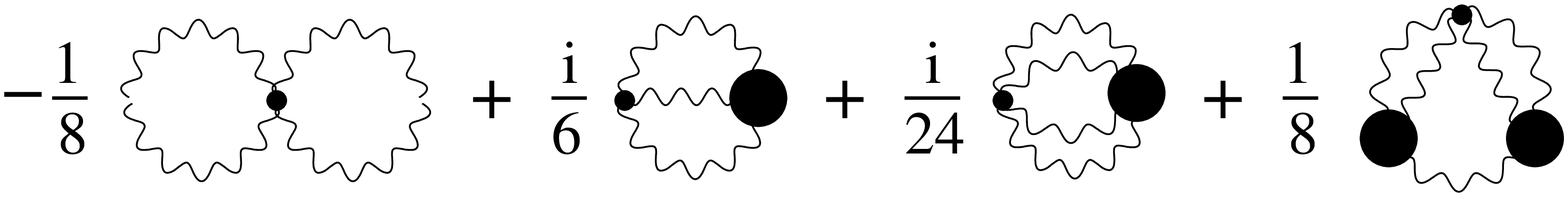,width=12.5cm}
\end{center}
\vspace*{-0.5cm}
\caption{\small The figure shows together with Fig.~\ref{fig:ym3loop2}
the diagrammatic representation of 
$\Gamma_2^0[D,G,V_{3},V_3^{\rm (gh)},V_{4}]$ as given
in Eq.~(\ref{eq:G20}). Here the wiggled lines denote the
gauge field propagator $D$ and the unwiggled lines the
ghost propagator $G$. The thick circles denote the dressed
and the small ones the classical vertices. This functional 
contains all terms of $\Gamma_2$ that depend on the classical 
vertices $g V_{03}$, $g V_{03}^{\rm (gh)}$ and $g^2 V_{04}$ for an 
$SU(N)$ gauge theory. There are no further contributions to 
$\Gamma_2^0$ appearing at higher order in the expansion. 
For the gauge theory with 
fermions there is in addition the same contribution as in
Fig.~\ref{fig:ym3loop2} with the unwiggled propagator lines
representing the fermion propagator $\Delta$ and the ghost vertices 
replaced by the corresponding fermion vertices $V_{03}^{\rm (f)}$ 
and $V_{3}^{\rm (f)}$ (cf.~Eq.~(\ref{eq:fermvertex})).}
\label{fig:ym3loop1}
\end{figure}
For the effective action to $\mathcal{O}(g^6)$ we obtain:  
\bea
\Gamma[D,G,V_{3},V_3^{\rm (gh)},V_{4}] &=& \frac{i}{2} \Tr \ln D^{-1}
+ \frac{i}{2} \Tr\, D_0^{-1} D 
- i \Tr \ln G^{-1}
- i \Tr\, G_0^{-1} G 
\nonumber\\
&& + \Gamma_2[D,G,V_{3},V_3^{\rm (gh)},V_{4}] \, ,
\label{eq:exact4PI}
\eea
with
\bea
\Gamma_2[D,G,V_{3},V_3^{\rm (gh)},V_{4}] 
&=& \Gamma_2^0[D,G,V_{3},V_3^{\rm (gh)},V_{4}]
+ \Gamma_2^{\rm int}[D,G,V_{3},V_3^{\rm (gh)},V_{4}] \, , 
\nonumber\\[0.2cm]
\Gamma_2^0[D,G,V_{3},V_3^{\rm (gh)},V_{4}] &=&
- \frac{1}{8} g^2 D^2 V_{04}
+ \frac{i}{6} g^2 D^3 V_{3} V_{03}
- i g^2 D G^2 V_3^{\rm (gh)} V_{03}^{\rm (gh)}
\nonumber\\
&& + \frac{i}{24} g^4 D^4 V_{4} V_{04}
+ \frac{1}{8} g^4 D^5 V_{3}^2 V_{04} \, , 
\label{eq:G20} \\[0.2cm]
\Gamma_2^{\rm int}[D,G,V_{3},V_3^{\rm (gh)},V_{4}] &=& 
- \frac{i}{12} g^2 D^3 V_{3}^2 
+ \frac{i}{2} g^2 D G^2 V_{3}^{\rm (gh)\, 2}
- \frac{i}{48} g^4 D^4 V_{4}^2  
\nonumber\\
&& - \frac{i}{24} g^4 D^6 V_{3}^4 
+ \frac{i}{3} g^4 D^3 G^3\, V_{3}^{\rm (gh)\, 3}V_{3} 
\nonumber\\
&& + \frac{i}{4} g^4 D^2 G^4\, V_{3}^{\rm (gh)\, 4}
+ \mathcal{O}(g^6) \, .
\label{eq:gamma2}
\eea
The contributions are displayed diagrammatically 
in Figs.~\ref{fig:ym3loop1} and~\ref{fig:ym3loop2} 
for $\Gamma_2^0$, and in Figs.~\ref{fig:ym3loop3} 
and~\ref{fig:ym3loop4} for $\Gamma_2^{\rm int}$.

The equivalence of the 4PI effective action to three-loop order with
the 3PI and $n$PI effective actions for $n \ge 5$ in the absence of 
sources follows along the lines of 
Sec.~\ref{sec:selfcomplete}. As a consequence, to three-loop order the 
$n$PI effective
action does not depend on higher vertices $V_5$, $V_6$, \ldots $V_n$.
In particular with vanishing sources the four-vertex is given by the 
classical one: 
\beq
\frac{\delta \Gamma^{\rm (3loop)}[D,G,V_3,V_3^{\rm (gh)},V_4]}{\delta V_4} = 
\frac{\delta
\Gamma^{\rm (3loop)}_2[D,G,V_3,V_3^{\rm (gh)},V_4]}{\delta V_4} = 0 
\,\, \Rightarrow \,\, V_4 = V_{04} .
\label{eq:classical4}
\eeq
If one plugs this into (\ref{eq:G20}) and (\ref{eq:gamma2}) one
obtains the three-loop 3PI effective action,
$\Gamma^{\rm (3loop)}[\phi,D,V_3,V_3^{\rm (gh)}]$. Similarly, to two-loop
order one has 
\bea
\frac{\delta \Gamma^{\rm (2loop)}[D,G,V_3,V_3^{\rm (gh)}]}{\delta V_3} 
&\!=\!& 
\frac{\delta \Gamma^{\rm (2loop)}_2[D,G,V_3,V_3^{\rm (gh)}]}{\delta V_3} = 0 
\,\, \Rightarrow \,\, V_3 = V_{03} \, , \nonumber\\
\frac{\delta 
\Gamma^{\rm (2loop)}[D,G,V_3,V_3^{\rm (gh)}]}{\delta V_3^{\rm (gh)}} 
&\!=\!& \frac{\delta \Gamma^{\rm (2loop)}_2[D,G,V_3,V_3^{\rm (gh)}]}
{\delta V_3^{\rm (gh)}} = 0 
\,\, \Rightarrow \,\, V_3^{\rm (gh)} = V_{03}^{\rm (gh)} \, ,
\nonumber
\eea
and equivalently for the fermion vertex $V_3^{\rm (f)}$.
To this order, therefore, the combinatorial factors of the
two-loop diagrams of Fig.~\ref{fig:ym3loop1} and
\ref{fig:ym3loop3} for the gauge part, as well as 
of Fig.~\ref{fig:ym3loop2} and
\ref{fig:ym3loop4} for the ghost/fermion part,
combine to give the result (\ref{eq:3loop2PISUN}) to two-loop order for
the 2PI effective action. 

\section{Equations of motion} 
\label{sec:equationsofm}

\begin{figure}[t]
\begin{center}
\epsfig{file=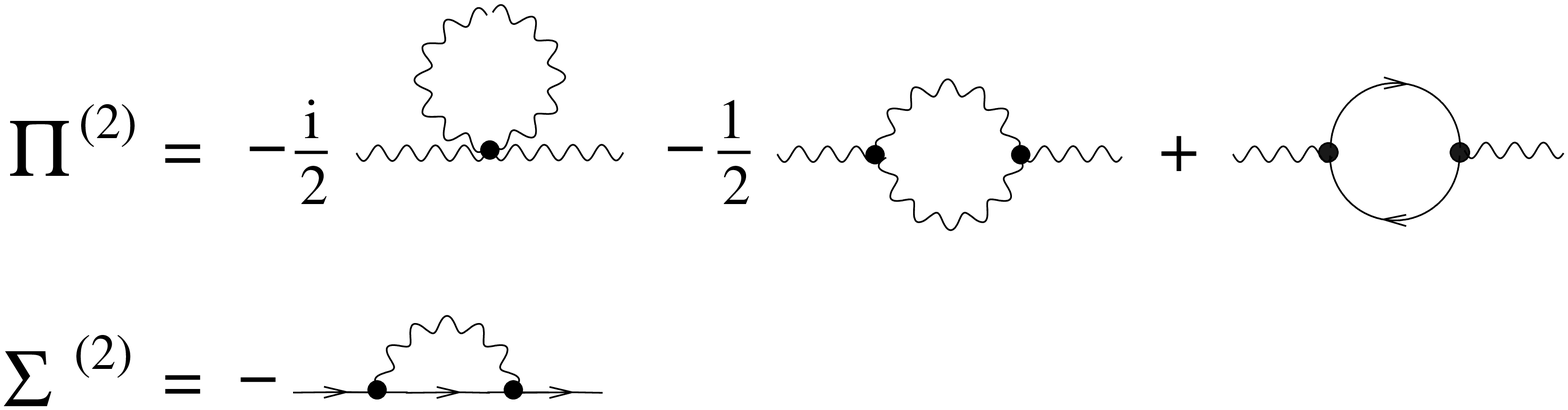,width=13.cm}
\end{center}
\vspace*{-0.5cm}
\caption{\small The self-energy for the gauge field ($\Pi$) and
the ghost/fermion ($\Sigma$) propagators as obtained 
from the self-consistently complete 
two-loop approximation of the effective action. 
Note that at this order all vertices correspond to 
the classical ones.}
\label{fig:ymmerge2}
\end{figure}

\begin{figure}[t]
\begin{center}
\epsfig{file=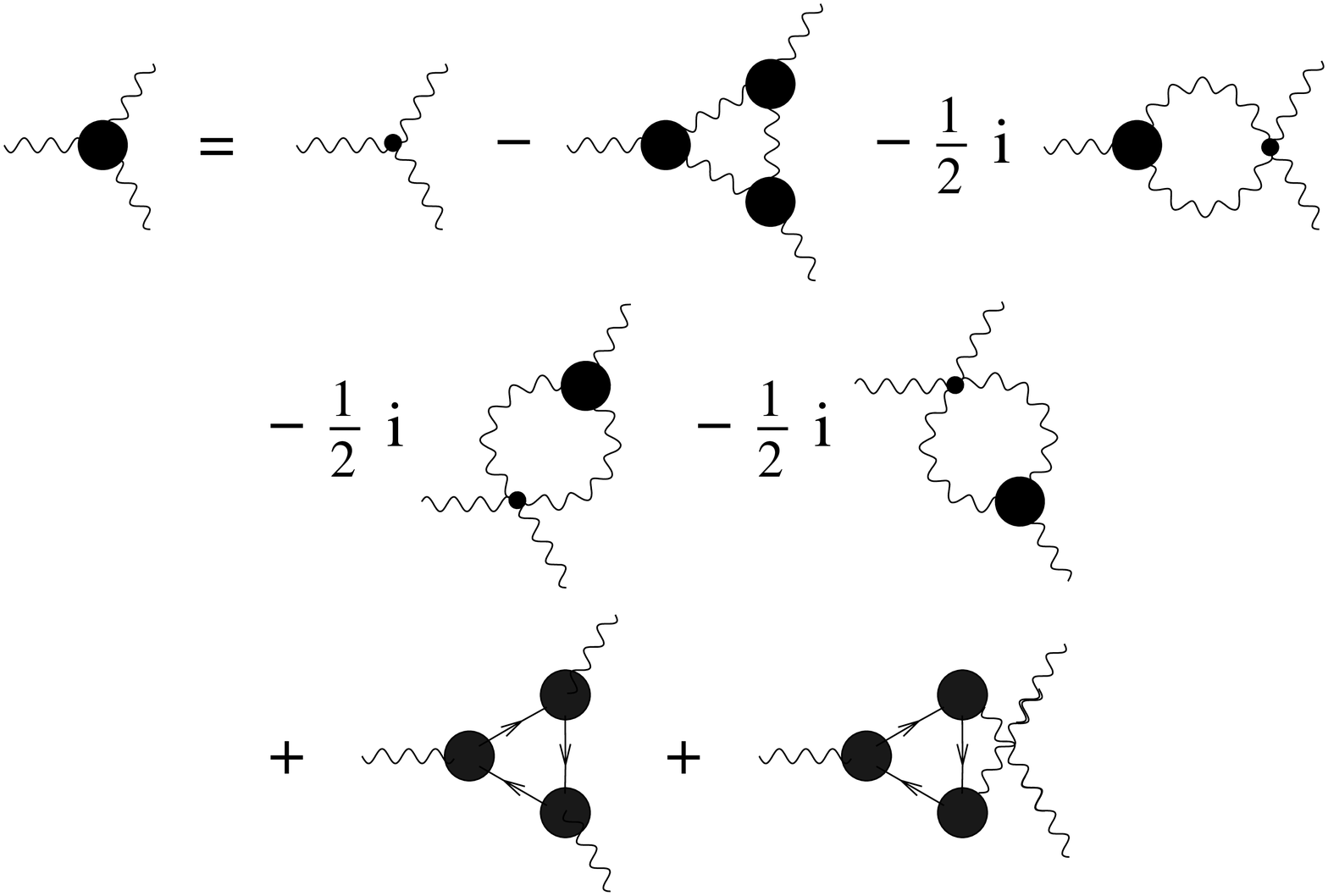,width=9.6cm}
\epsfig{file=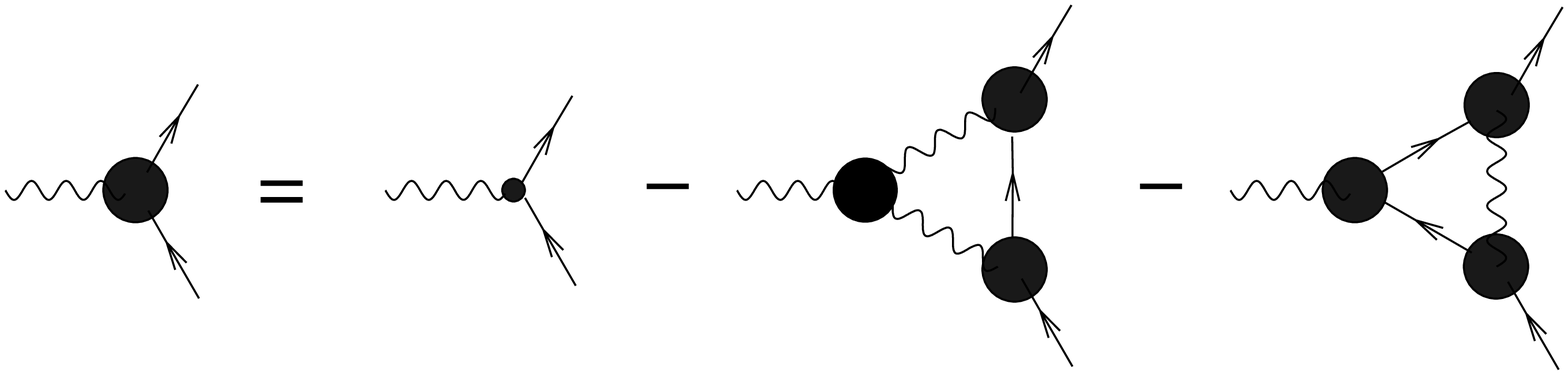,width=9.6cm}
\end{center}
\vspace*{-0.5cm}
\caption{\small The gauge field three-vertex as well as the ghost (fermion)
vertex as obtained from the self-consistently complete 
three-loop approximation of the effective action.
Note that apart from the isolated classical three-vertex, all 
vertices in the equations correspond to dressed ones since at
this order the four-vertex equals the classical vertex.}
\label{fig:ymver3}
\end{figure}

In the last section we have seen that to two-loop order 
the proper vertices of the $n$PI effective action 
correspond to the classical ones. Accordingly, at this order
the only non-trivial equations of motion in the absence
of background fields are those for the two-point functions:
\beq
\frac{\delta \Gamma}{\delta D} = 0 \, , \quad \, 
\frac{\delta \Gamma}{\delta G} = 0 \, , \quad \,
\frac{\delta \Gamma}{\delta \Delta} = 0\, ,
\label{eq:statprop}
\eeq
for vanishing sources. Applied to an $n$PI effective
action ($n>1$), as e.g.~(\ref{eq:exact4PI}),
one finds for the gauge field propagator:
\beq
D^{-1} = D^{-1}_0 
- \Pi \, ,
\label{eq:SDforD}  
\eeq
where the proper self-energy is given by
\beq
\Pi = 2 i\, 
\frac{\delta \Gamma_2}{\delta D}  \, .
\eeq 
The ghost propagator and self-energy are 
\beq
G^{-1} = G^{-1}_0 
- \Sigma \quad , \quad 
\Sigma = -i \frac{\delta \Gamma_2}{\delta G} \, ,
\label{eq:SDforG}  
\eeq
and equivalently for the fermion propagator $\Delta$.
The self-energies to this order are shown in diagrammatic form 
in Fig.~\ref{fig:ymmerge2}. In contrast, for the three-loop 
effective action the three-vertices get dressed and the stationarity 
conditions, 
\beq 
\frac{\delta \Gamma}{\delta V_{3}} = 0 \, , \quad \, 
\frac{\delta \Gamma}{\delta V_{3}^{\rm (gh)}} = 0 \, , \quad \,
\frac{\delta \Gamma}{\delta V_{3}^{\rm (f)}} = 0 \, , 
\label{eq:statver}
\eeq
applied to (\ref{eq:exact4PI})--(\ref{eq:gamma2})
lead to the equations shown in Fig.~\ref{fig:ymver3}.
Here the diagrammatic form of the contributions is always 
the same for the ghost and for the fermion propagators or vertices. We 
therefore only give the expressions for the gauge-ghost
system. If fermions are present, the respective diagrams
have to be added in a straightforward way.  
\begin{figure}[t]
\begin{center}
\epsfig{file=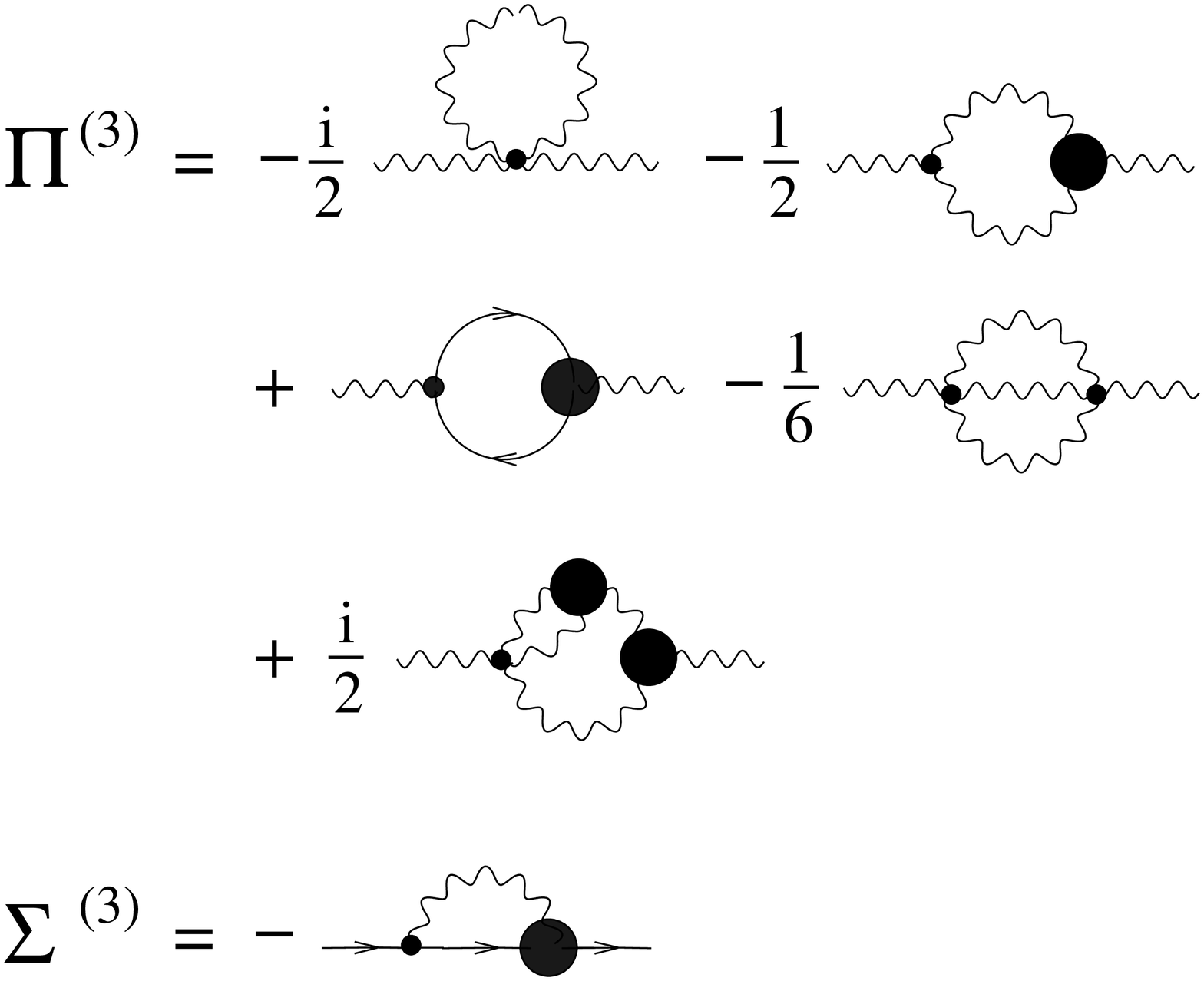,width=9.3cm}
\end{center}
\vspace*{-0.5cm}
\caption{\small The self-energy for the gauge field ($\Pi$) and
the ghost/fermion ($\Sigma$) propagators as obtained 
from the self-consistently complete 
three-loop approximation of the effective action.
(Cf.~Fig.~\ref{fig:ymver3} for the vertices.)}
\label{fig:ymmerge3}
\end{figure}

The respective self-energies to this order are displayed 
in Fig.~\ref{fig:ymmerge3}. It should be emphasized that 
their relatively simple form is a consequence of the
equations for the proper vertices, Fig.~\ref{fig:ymver3}.  
To see this we consider first the many terms generated 
by the functional derivative
of (\ref{eq:G20}) and (\ref{eq:gamma2}) with respect to
the gauge field propagator:
\bea
\Pi^{(3)}\, \equiv\, 2i \frac{\delta \Gamma_2^{\rm (3loop)}}{\delta D} &\!=\!&
- \frac{i}{2}\, 
\begin{minipage}{1.2cm}
\epsfig{file=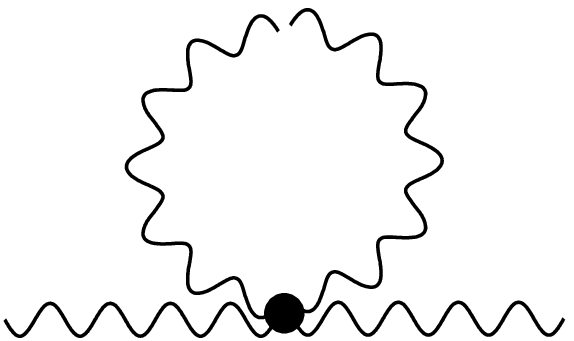,width=1.2cm}
\end{minipage}
- 
\begin{minipage}{1.4cm}
\epsfig{file=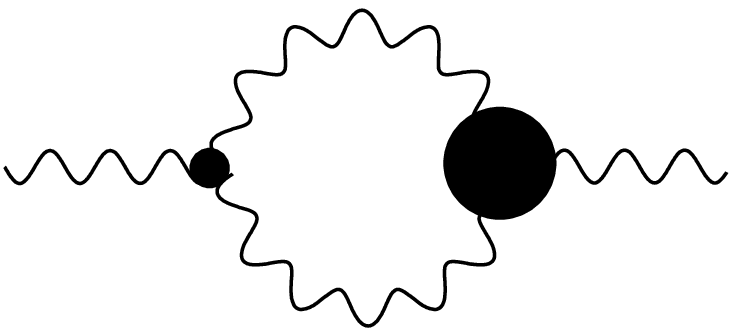,width=1.4cm}
\end{minipage}
+ \frac{1}{2}\, 
\begin{minipage}{1.4cm}
\epsfig{file=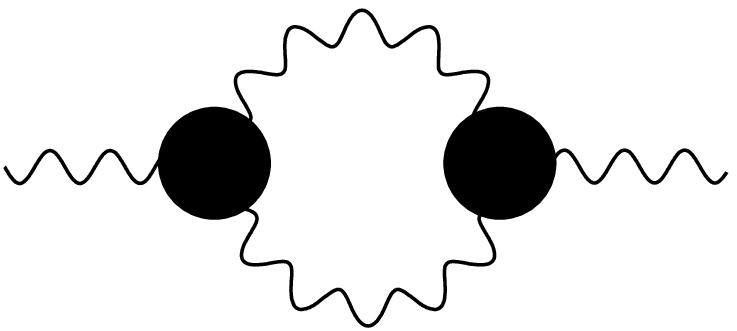,width=1.4cm}
\end{minipage}
+ 2\, 
\begin{minipage}{1.4cm}
\epsfig{file=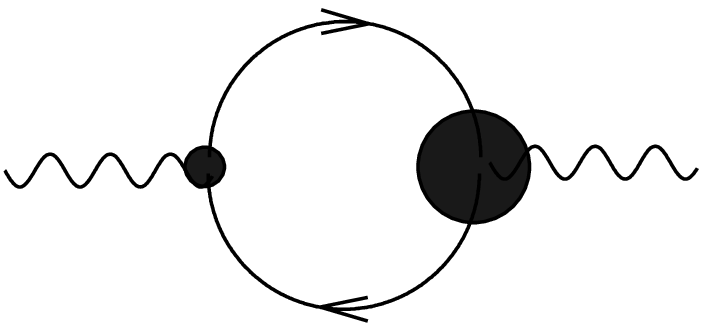,width=1.4cm}
\end{minipage}
\nonumber\\
&&
-\,  
\begin{minipage}{1.4cm}
\epsfig{file=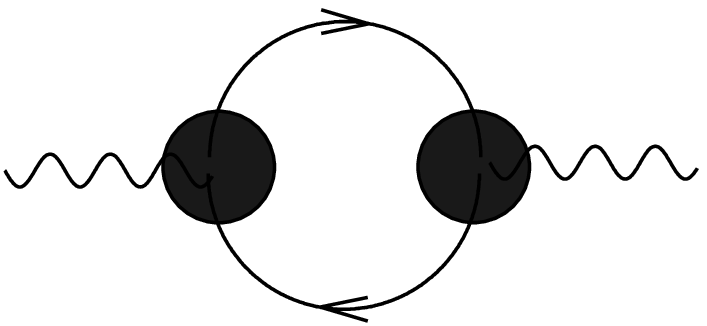,width=1.4cm}
\end{minipage}
- \frac{1}{3}\, 
\begin{minipage}{1.4cm}
\epsfig{file=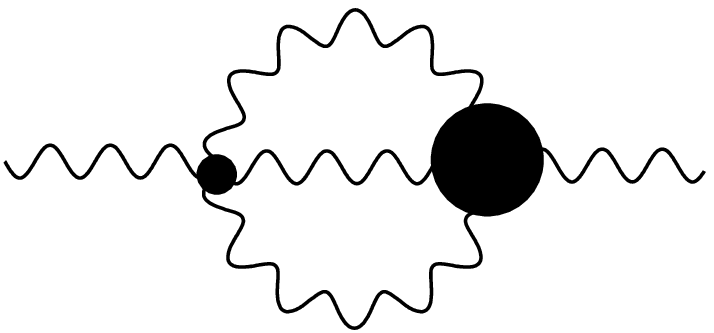,width=1.4cm}
\end{minipage}
+ \frac{1}{6}\, 
\begin{minipage}{1.4cm}
\epsfig{file=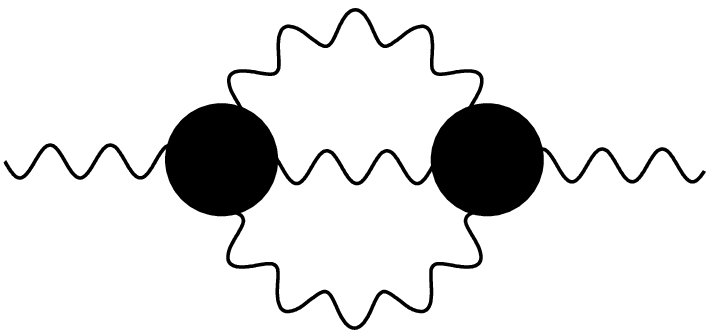,width=1.4cm}
\end{minipage}
+ i\, 
\begin{minipage}{1.4cm}
\epsfig{file=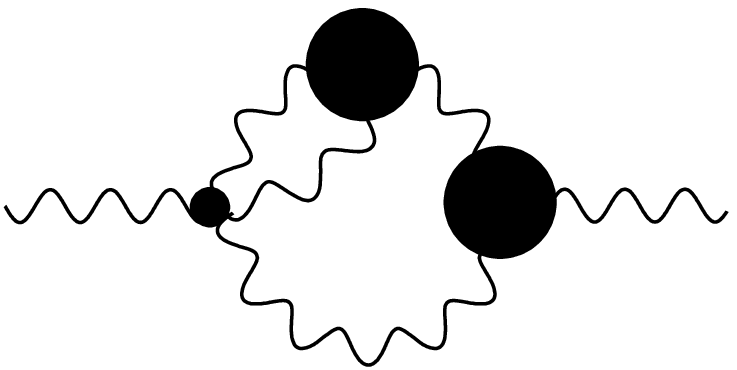,width=1.4cm}
\end{minipage}
\\
&& 
+ \frac{i}{4}\, 
\begin{minipage}{1.7cm}
\epsfig{file=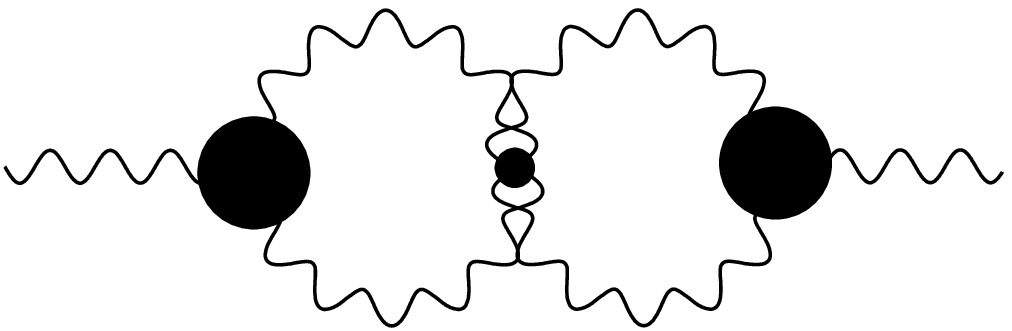,width=1.7cm}
\end{minipage}
+ \frac{1}{2}\, 
\begin{minipage}{1.6cm}
\epsfig{file=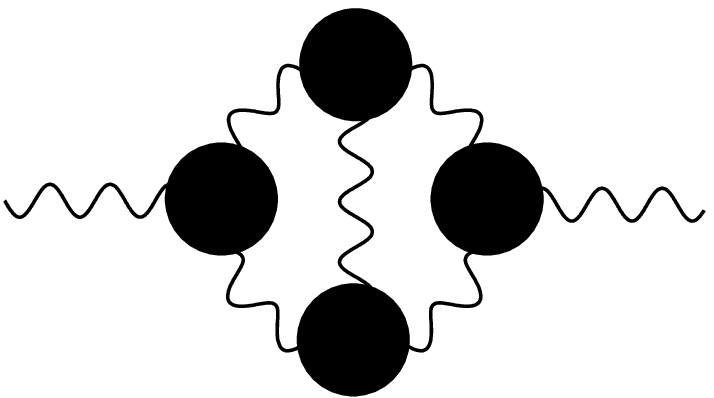,width=1.6cm}
\end{minipage}
- 2\, 
\begin{minipage}{1.6cm}
\epsfig{file=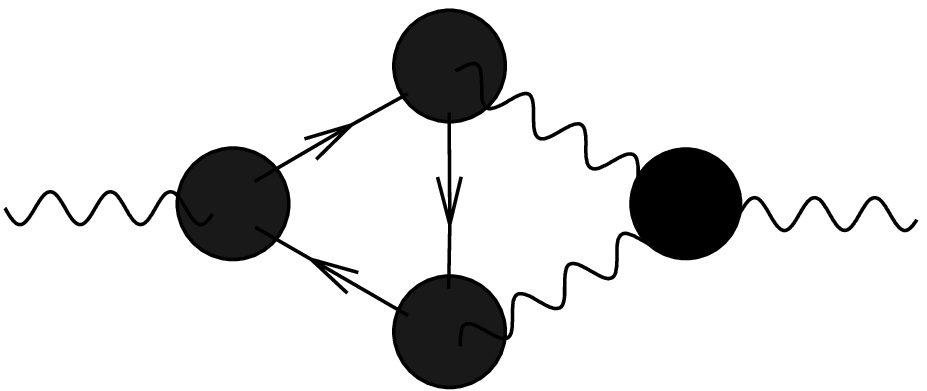,width=1.6cm}
\end{minipage}
-\, 
\begin{minipage}{1.6cm}
\epsfig{file=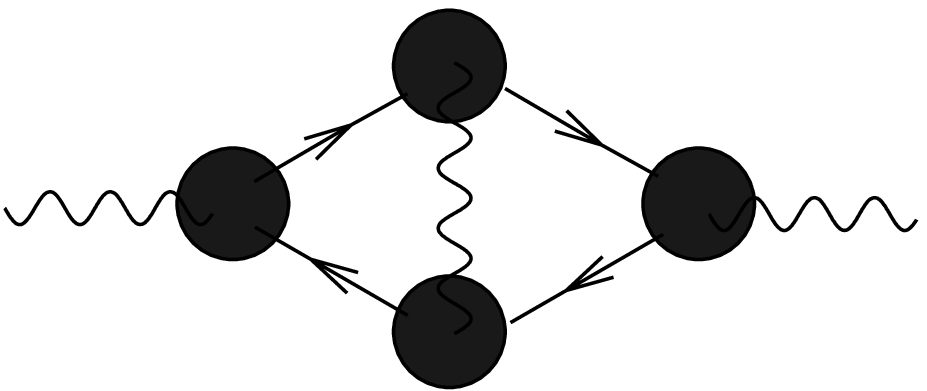,width=1.6cm}
\end{minipage} \, .
\nonumber
\eea
The short form for the self-energy of Fig.~\ref{fig:ymmerge3}
is obtained through cancellations by replacing in the above 
expression 
\bea
 \frac{1}{2}\, 
\begin{minipage}{1.4cm}
\epsfig{file=ymsun2.eps,width=1.4cm}
\end{minipage}
&\!=\!& \frac{1}{2}\, 
\begin{minipage}{1.4cm}
\epsfig{file=ymsun.eps,width=1.4cm}
\end{minipage}
-\frac{1}{2}\, 
\begin{minipage}{1.6cm}
\epsfig{file=ymdum3.eps,width=1.6cm}
\end{minipage}
-\frac{i}{4}\, 
\begin{minipage}{1.6cm}
\epsfig{file=ymdum1.eps,width=1.6cm}
\end{minipage}
\nonumber\\
&&
-\frac{i}{2}\, 
\begin{minipage}{1.6cm}
\epsfig{file=ymshade.eps,width=1.6cm}
\end{minipage}
+\, 
\begin{minipage}{1.6cm}
\epsfig{file=ymdum2.eps,width=1.6cm}
\end{minipage}
\eea
as well as
\bea
-\, 
\begin{minipage}{1.4cm}
\epsfig{file=ymsunghost2.eps,width=1.4cm}
\end{minipage}
&\!=\!& -\, 
\begin{minipage}{1.4cm}
\epsfig{file=ymsunghost.eps,width=1.4cm}
\end{minipage}
+\, 
\begin{minipage}{1.6cm}
\epsfig{file=ymdum4.eps,width=1.6cm}
\end{minipage}
+\, 
\begin{minipage}{1.6cm}
\epsfig{file=ymdum2.eps,width=1.6cm}
\end{minipage} \, .
\eea
The latter equations follow from inserting 
the expressions for the dressed vertices
of Fig.~\ref{fig:ymver3}. Noting in addition that 
the proper four-vertex to this order 
corresponds to the classical one 
(cf.~(\ref{eq:classical4})) leads
to the result. Along the very same lines a 
similar cancellation
yields the compact form of the 
ghost/fermion self-energy displayed in 
Fig.~\ref{fig:ymmerge3}.

\subsection{Comparison with Schwinger-Dyson equations}
\label{sec:compSD}

The equations of motions of the last section are self-consistently
complete to two-loop/three-loop order 
of the $n$PI effective action for arbitrarily large $n$.
We now compare them with conventional Schwinger-Dyson (SD) equations,
which represent identities between $n$-point functions.
Clearly, without approximations the equations of motion
obtained from an exact \mbox{$n$PI} effective action and
the exact (SD) equations have to agree since one can
always map identities onto each other. However, in general
this is no longer the case for a given order in the
loop expansion of the $n$PI effective action.

By construction each diagram in a SD equation contains
at least one classical vertex~\cite{SD}. In general, this is not the case
for equations obtained from the $n$PI effective action:
The loop contributions of $\Gamma_2^{\rm int}$ in Eq.~(\ref{eq:gamma2})
or Figs.~\ref{fig:ym3loop3}--\ref{fig:ym3loop4} are 
solely expressed in terms of full vertices.
However, to a given loop-order cancellations can occur for
those diagrams in the equations of motion which do not
contain a classical vertex. For the three-loop effective
action result this has been demonstrated in Sec.~\ref{sec:equationsofm}
for the two-point functions. Indeed, the equations for the 
two-point functions shown in Fig.~\ref{fig:ymmerge3} correspond
to the SD equations, if one takes into account that to the considered 
order the four-vertex is trivial and given by the classical one 
(cf.~\ref{eq:classical4}). However, such a correspondence
is not true for the proper three-vertex to that order.

\begin{figure}[t]
\begin{center}
\epsfig{file=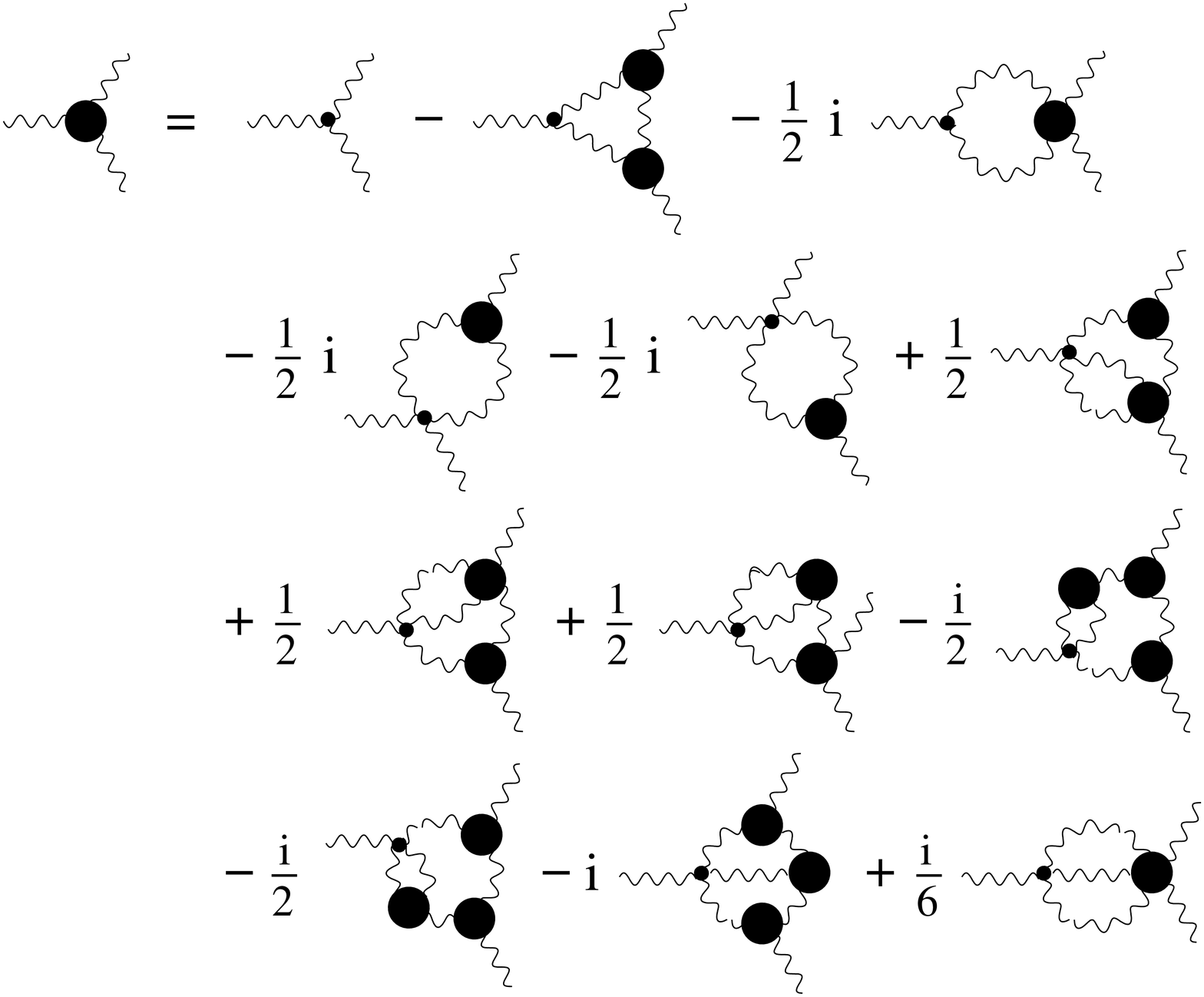,width=10.3cm}
\end{center}
\vspace*{-0.5cm}
\caption{\small Standard
Schwinger-Dyson equation for the proper three-vertex $V_3$.
We have not displayed additional diagrams involving ghost or
fermion vertices for brevity.
We show it for comparison with the three-loop effective action result
displayed in Fig.~\ref{fig:ymver3}. One observes that a naive
truncation of the Schwinger-Dyson equation at the one-loop level
does not agree with the latter, since the second and third diagram contain a 
classical three-vertex instead of a dressed one as in 
Fig.~\ref{fig:ymver3}. (Note that the four-vertex equals the
classical one at this order in the self-consistently complete
loop expansion.)}
\label{fig:SDver3}
\end{figure}

As an example, we show in 
Fig.~\ref{fig:SDver3} the
standard (SD) equation for the proper three-vertex,
where we neglect for a moment the additional diagrams
coming from ghost/fermion degrees of 
freedom (cf.~e.g.~\cite{Kajantie:2001hv}). 
One finds that a naive neglection of the two-loop contributions
of that equation would not lead to the effective action result 
for the three-vertex shown in Fig.~\ref{fig:ymver3}.
Of course, the straightforward one-loop truncation of the 
SD equation would not even respect the property of 
$V_3$ being completely symmetric in its space-time and group
labels. This is the well-known problem of loop-expansions
of SD equations, where one encounters the ambiguity of whether
classical or dressed vertices should be employed at 
a given truncation order.

We emphasize that these problems are absent using effective action 
techniques. The fact that all equations of motion are obtained from
the same approximation of the effective action puts stringent 
conditions on their form. More precisely, a self-consistently
complete approximation has the property that the order of 
differentiation of, say, $\Gamma[D,V]$ with respect to the
propagator $D$ or the vertex $V$ does not affect the equations of
motion. Consider for instance: 
\beq
\frac{\delta \Gamma[D,V=V(D)]}{\delta D}
= \frac{\delta \Gamma}{\delta D}\Big|_{V}
+ \frac{\delta \Gamma}{\delta V}\Big|_{D}
\, \frac{\delta V}{\delta D}  \, .
\eeq
If $V = V(D)$ is the result of the stationary condition
$\delta \Gamma/\delta V = 0$ then the above 
corresponds to the correct stationarity condition for
the propagator for fixed $V$: $\delta \Gamma/\delta D = 0$. 
In contrast, with some ansatz $V=f(D)$ that does not
fulfill the stationarity condition of the 
effective action, the equation of motion for the
propagator would receive additional corrections
$\sim \delta V/\delta D$. In particular, it would be
inconsistent to use the equation of motion for the
propagator $\delta \Gamma/\delta D = 0$
(cf.~e.g.~Fig.~\ref{fig:ymmerge3} which corresponds 
to the SD equation result) but not the equation $\delta \Gamma/\delta V = 0$
for the vertex (cf.~Fig.~\ref{fig:ymver3}).

In turn, one can conclude that a wide class of employed truncations 
of exact SD equations cannot be obtained from the $n$PI effective 
action: this concerns those approximations which use the
exact SD equation for the propagator but make an ansatz
for the vertices that differs from the one displayed in 
Fig.~\ref{fig:ymver3}. The differences are, however, typically
higher order in the perturbative coupling expansion and there
may be many cases, in particular in vacuum or thermal equilibrium,
where some ansatz for the vertices is a very efficient way to  
proceed. Out of equilibrium however, as mentioned above, 
the ``conserving'' property 
of the effective action approximations
can have important consequences, since the 
effective loss of initial conditions and the presence of basic constants 
of motion such as energy conservation is crucial.

\section{Nonequilibrium evolution equations}
\label{sec:nonequilibrium}

The above equations of motion have the form of 
self-consistent or ``gap'' equations, as in 
(\ref{eq:SDforD}) or (\ref{eq:SDforG}), which is very suitable  
for vacuum or thermal equilibrium problems. In this case the
time integrations displayed in Sec.~\ref{sec:nonabgauge}
run along the real axis ($\int_x \equiv \int_{-\infty}^{\infty}
{\rm d}^{d+1} x$) or along the imaginary time-axis 
($\int_x \equiv \int_{0}^{- i \beta} {\rm d} x^0
\int {\rm d} {\bf x}$) up to the
inverse temperature $\beta$, respectively~\cite{Kapusta}.
For nonequilibrium time-evolution problems it is useful
to rewrite the equations in a standard way such that they are suitable for
initial-value problems. The time integration in this case
starts at some initial time and involves a closed path $\mathcal C$
along the real axis ($\int_x \equiv \int_{\mathcal C} {\rm d} x^0
\int {\rm d} {\bf x}$)~\cite{Schwinger:1961qe}.\footnote{Here we will consider
Gaussian initial conditions, which represents
no approximation but restricts the class of initial conditions.
For details see~e.g.~Refs.~\cite{Berges:2000ur,Berges:2001fi}.}    

Up to $\mathcal{O}(g^6)$ corrections in the self-consistently
complete expansion of the effective action,
the four-vertex parametrizing the diagrams of 
Figs.~\ref{fig:ymmerge3}--\ref{fig:ymver3} corresponds to the classical vertex.
At this order of approximation there is, therefore, no
distinction between the coupling expansion of the 3PI and 4PI 
effective action. To discuss the relevant differences between
the 2PI and 3PI expansions for time evolution problems, we will
use the language of QED for simplicity, where no four-vertex
appears. However, the evolution equations of this section can be
straightforwardly transcribed to the nonabelian case by 
taking into account in addition to the equation for the gauge--fermion 
three-vertex those for the gauge--ghost and gauge three-vertex
(cf.~Fig.~\ref{fig:ymver3}).        
In the following the effective action is a functional of the gauge
field propagator $D_{\mu\nu}(x,y)$, the fermion propagator 
$\Delta(x,y)$ and the gauge-fermion vertex 
$V_{3\, \mu}^{\rm (f)}(x,y;z)$, where we suppress Dirac
indices and we will write
$V_{3}^{\rm (f)} \equiv V$. According to 
Eqs.~(\ref{eq:exact4PI})---(\ref{eq:gamma2}) one has in this case
\beq
\Gamma_2[D,\Delta,V] 
= \Gamma_2^0[D,\Delta,V]
+ \Gamma_2^{\rm int}[D,\Delta,V] \, , 
\eeq
with
\beq
\Gamma_2^0 = - i g^2 \int_{x y z u} \tr\left[ \gamma_\mu
\Delta(x,y) V_{\nu}(y,z;u) \Delta(z,x) D^{\mu\nu}(x,u) \right] \, ,
\label{eq:gamma2qed}
\eeq
where the trace acts in Dirac space. For the given order of
approximation there are two 
distinct contributions to $\Gamma_2^{\rm int}$:
\bea
\Gamma_2^{\rm int}&=&\Gamma_2^{(a)} + \Gamma_2^{(b)} 
+ \mathcal{O}\left(g^6\right)\, ,
\label{eq:gamma2qedint}\\
\Gamma_2^{(a)} &=&  \frac{i}{2} g^2 \int_{x y z u v w}
\tr\left[ V_{\mu}(x,y;z)
\Delta(y,u) V_{\nu}(u,v;w) \Delta(v,x) D^{\mu\nu}(z,w) \right] \, ,
\nonumber\\
\Gamma_2^{(b)} &=&  \frac{i}{4} g^4 \int_{x y z u v w x' y' z' u' v' w'}
\tr\left[ V_{\mu}(x,y;z)
\Delta(y,u) V_{\nu}(u,v;w) \Delta(v,x') 
 \right. \nonumber\\
&& \left. V_{\rho}(x',y';z') \Delta(y',u') V_{\sigma}(u',v';w') 
\Delta(v',x) D^{\mu\rho}(z,z') D^{\nu\sigma}(w,w') \right] \, .
\nonumber
\eea
The equations of motions for the propagators and vertex 
are obtained from the stationarity conditions (\ref{eq:statprop})
and (\ref{eq:statver}) for the effective action. 
To convert (\ref{eq:SDforD}) for the photon propagator 
into an equation which is more suitable 
for initial value problems, we convolute with $D$ from the right 
and obtain for the considered case of vanishing `background' fields, 
e.g.~for covariant gauges:
\bea
\left[{g^\mu}_\gamma \square - (1-\xi^{-1}) \partial^\mu 
\partial_{\gamma} \right]_x D^{\gamma\nu}(x,y)
- i \int_z {\Pi^\mu}_{\gamma}(x,z) D^{\gamma\nu}(z,y) 
\nonumber \\
 = i g^{\mu\nu} \delta_\mathcal{C}(x-y)  \, .
\label{eq:evolD}
\eea 
Similarly, the corresponding equation of (\ref{eq:SDforG}) yields 
the evolution equation for the fermion propagator: 
\beq
i \partial_x \!\slash \, \Delta(x,y) - i \int_z \Sigma(x,z) \Delta(z,y)
= i \delta_{\C}(x-y) \, .
\label{eq:evolDel}
\eeq 
Using the results of Sec.~\ref{sec:equationsofm} the self-energies
are
\bea
\Sigma(x,y) &=& - g^2 \int_{z'z''} D_{\mu\nu}(z',y)
V^{\mu}(x,z'';z') \Delta(z'',y) \gamma^{\nu} \, ,
\label{eq:selffermexact}\\
\Pi^{\mu\nu}(x,y) &=& g^2 \int_{z'z''} \tr\, \gamma^{\mu}
\Delta(x,z') V^{\nu}(z',z'';y) \Delta(z'',x)  \, .
\label{eq:selfexact}
\eea
Note that the form of the self-energies is exact for known three-vertex.
To see this within the current framework, we note that the self-energies can
be expressed in terms of $\Gamma_2^0$ only. The latter receives no further 
corrections at higher order in the expansion (cf.~Sec.~\ref{sec:calcSUN}), 
and thus the expression is exactly known: With
\bea
\int_z \Sigma(x,z) \Delta(z,y)
= - i \int_z \left( \frac{\delta \Gamma_2^0}{\delta \Delta(z,x)} 
+ \frac{\delta \Gamma_2^{\rm int}}{\delta \Delta (z,x)} \right) \Delta(z,y) 
\, ,
\eea
and since $\Gamma_2^{\rm int}$ is only a functional of $V \Delta D^{1/2}$
(cf.~Sec.~\ref{sec:calcSUN}) one can use the identity
\bea
\int_z \frac{\delta \Gamma_2^{\rm int}}{\delta \Delta(z,x)} \Delta(z,y)
&=&  \int_{z z'} V_{\mu}(x,z;z')  
\frac{\delta \Gamma_2^{\rm int}}{V_{\mu}(y,z;z')} \nonumber\\
&=& - \int_{z z'} V_{\mu}(x,z;z')  
\frac{\delta \Gamma_2^0}{V_{\mu}(y,z;z')}
\label{eq:delverrew}
\eea
to express everything in terms of the known\footnote{This
can also be directly verified from (\ref{eq:gamma2qed}) 
to the given order of approximation.} $\Gamma_2^0$.
The last equality in (\ref{eq:delverrew}) uses that
$\delta (\Gamma_2^0+\Gamma_2^{\rm int})/\delta \Delta = 0$.  
A similar discussion can be done for the photon self-energy.
As a consequence,
all approximations are encoded in the equation for the vertex, 
which is obtained from (\ref{eq:gamma2qedint}) as: 
\bea
V^{\mu} (x,y;z) &=& V_0^{\mu} (x,y;z)  
- g^2 \int_{v w x' y' u' w'}
V_{\nu} (x,v;w) \Delta(v,x')V^{\mu} (x',y';z)
 \nonumber\\
&& \Delta(y',u') V_{\sigma} (u',y;w') D^{\sigma\nu}(w',w) 
+ \mathcal{O}\left(g^4\right)\, ,
\label{eq:vertexeom}
\eea
where
\beq
V_0^{\mu} (x,y;z) = \gamma^{\mu} \delta(x-z) \delta(z-y) \, . 
\eeq
For the self-consistently complete two-loop approximation
the self-energies are given by  
\bea
\Sigma(x,y) &=& - g^2 D_{\mu\nu}(x,y)
\gamma^{\mu} \Delta(x,y) \gamma^{\nu} + \mathcal{O}\left(g^4\right) 
\, ,\\[0.2cm]
\Pi^{\mu\nu}(x,y) &=& g^2\, \tr\, \gamma^{\mu}
\Delta(x,y) \gamma^{\nu} \Delta(y,x)  + \mathcal{O}\left(g^4\right)\, .
\label{eq:selfg2}
\eea

\subsection{Spectral and statistical correlation functions}
\label{sec:frho}

We decompose the two-point functions into `spectral' and 
`statistical components' by writing~\cite{Aarts:2001qa,Berges:2001fi}
\bea
D^{\mu\nu}(x,y) 
&=& F_D(x,y)^{\mu\nu} - \frac{i}{2} \rho_D(x,y)^{\mu\nu} 
\, {\rm sign} (x^0-y^0) \, .
\eea
Here $\rho_D$ corresponds to the spectral function 
and $F_D$ is the so-called statistical two-point 
function\footnote{
Note that $\rho_D$ is determined by the commutator of two fields,
while $F_D$ by the anti-commutator. Out of equilibrium, where
the fluctuation dissipation theorem does not hold in general, both 
$F_D$ and $\rho_D$ are lin.~independent two-point functions. In terms
of the conventional decomposition
\bea
D^{\mu\nu}(x,y) = 
\Theta(x^0 - y^0) D_>(x,y)^{\mu\nu}
+ \Theta(y^0 - x^0) D_<(x,y)^{\mu\nu} \nonumber
\eea
one has
\bea
F_D(x,y)^{\mu\nu} = \frac{1}{2} 
\left( D_>(x,y)^{\mu\nu} + D_<(x,y)^{\mu\nu}
\right)\quad , \quad 
\rho_D(x,y)^{\mu\nu} = i \left( D_>(x,y)^{\mu\nu} - D_<(x,y)^{\mu\nu}
\right) \, . \nonumber
\eea
For Grassmann fields the spectral function corresponds to
the anti-commutator of two fields and the statistical two-point function
is determined by the commutator~\cite{Berges:2002wr}.
}.
Equivalently, the decomposition identity of the fermion two-point function into
spectral and statistical components reads~\cite{Berges:2002wr}
\bea
\Delta(x,y) 
&=& F_{\Delta}(x,y) 
- \frac{i}{2} \rho_{\Delta}(x,y)\, {\rm sign} (x^0-y^0) \, .
\eea
The same decomposition can be done for the corresponding
self-energies:\footnote{
If there is a local contribution to the proper self-energy,
we write 
\bea
\Sigma (x,y) = - i\, 
\Sigma^{\rm (local)}(x)\,\delta (x-y)
+ \Sigma^{\rm (nonlocal)}(x,y) \, ,\nonumber
\eea
and the decomposition (\ref{eq:sigmadec}) is taken for 
$\Sigma^{\rm (nonlocal)}(x,y)$. In this case the local contribution
gives rise to an effective space-time dependent fermion mass term
$\sim \Sigma^{\rm (local)}(x)$.}
\bea
\Pi^{\mu\nu}(x,y) 
&=& \Pi_{(F)}(x,y)^{\mu\nu} - \frac{i}{2} \Pi_{(\rho)}(x,y)^{\mu\nu} 
\, {\rm sign} (x^0-y^0) \, , \\
\Sigma(x,y) 
&=& \Sigma_{(F)}(x,y) 
- \frac{i}{2} \Sigma_{(\rho)} (x,y)\, {\rm sign} (x^0-y^0) \, .
\label{eq:sigmadec}
\eea
Since the above decomposition for the propagators and self-energies
makes the time-ordering explicit, we can evaluate 
the RHS of (\ref{eq:evolD}) along the time contour~\cite{Berges:2001fi},
and one finds the evolution 
equations~(cf.~also~\cite{Blaizot:2001nr}):
\bea
\left[{g^\mu}_\gamma \square - (1-\xi^{-1}) \partial^\mu 
\partial_{\gamma} \right]_x \rho_D(x,y)^{\gamma\nu}
&=& \int_{y^0}^{x^0} {\rm d}z\, 
\Pi_{(\rho)}(x,z)^{\mu \gamma} \rho_D{(z,y)_{\gamma}}^{\nu} \,\,\, ,
\nonumber\\
\label{eq:rho}\\
\left[{g^\mu}_\gamma \square - (1-\xi^{-1}) \partial^\mu 
\partial_{\gamma} \right]_x F_D(x,y)^{\gamma\nu}
&=& \int_{t_0}^{x^0} {\rm d}z\, 
\Pi_{(\rho)}(x,z)^{\mu \gamma} F_D{(z,y)_{\gamma}}^{\nu} 
\nonumber\\
&-& \int_{t_0}^{y^0} {\rm d}z\, 
\Pi_{(F)}(x,z)^{\mu \gamma} \rho_D{(z,y)_{\gamma}}^{\nu} \,\,\, ,
\nonumber\\
\label{eq:F}
\eea
where we used the abbreviated notation $\int_{t_1}^{t_2}
{\rm d}z \equiv \int_{t_1}^{t_2} {\rm d}z^0 
\int_{-\infty}^{\infty} {\rm d} \bf{z}$. The equations of
motion for the fermion spectral and statistical correlators
are obtained from (\ref{eq:evolDel})~\cite{Berges:2002wr}: 
\bea 
i {\partial\,\slash}_{\!x}  \rho_{\Delta} (x,y) &=& 
 \int_{y^0}^{x^0} {\rm d}z\,  \Sigma_{(\rho)} (x,z)\rho_{\Delta} (z,y) \, , 
\label{eq:rhoexact}\\
i {\partial\,\slash}_{\!x}  F_{\Delta}(x,y) &=& 
 \int_0^{x^0} {\rm d}z\, \Sigma_{(\rho)}(x,z) F_{\Delta}(z,y)
- \int_0^{y^0} {\rm d}z\, \Sigma_{(F)}(x,z)\rho_{\Delta}(z,y) \, .
\nonumber\\
\label{eq:Fexact}
\eea
For known self-energies the equations (\ref{eq:rho})--(\ref{eq:Fexact}) 
are exact. One observes that the form of their RHS is independent of 
whether it describes a boson or a fermion correlator.  

A similar discussion as for the two-point functions can also be done 
for the higher correlation functions. For the three-vertex 
we write
\beq
V^{\mu} (x,y;z) =  V_0^{\mu} (x,y;z) + \bar{V}^{\mu} (x,y;z)\, .
\label{eq:separation}
\eeq
and the corresponding decomposition into spectral and
statistical components reads
\bea
\lefteqn{ \bar{V}^{\mu} (x,y;z) = }
\nonumber\\
&& U_{(F)}(x,y;z)^{\mu}\, {\rm sign} (y^0-x^0)\, {\rm sign} (z^0-x^0) 
- \frac{i}{2} U_{(\rho)}(x,y;z)^{\mu}\, {\rm sign} (y^0-z^0)
\nonumber\\
&\!+\!\!& V_{(F)}(x,y;z)^{\mu}\, {\rm sign} (x^0-z^0)\, {\rm sign} (y^0-z^0) 
- \frac{i}{2} V_{(\rho)}(x,y;z)^{\mu}\, {\rm sign} (x^0-y^0)
\nonumber\\
&\!+\!\!& W_{(F)}(x,y;z)^{\mu}\, {\rm sign} (z^0-y^0)\, {\rm sign} (x^0-y^0) 
- \frac{i}{2} W_{(\rho)}(x,y;z)^{\mu}\, {\rm sign} (z^0-x^0) .
\nonumber\\
\label{eq:vbar}
\eea
This will be discussed further in the appendix.

\section{Kinetic theory and the LPM effect}
\label{sec:kinetic}

As an application we will consider the above equations 
in a standard ``on-shell'' limit which is typically 
employed in the literature
to derive kinetic equations for effective particle number
densities~\cite{Blaizot:2001nr}. We will see that 
since the lowest order contribution to the kinetic equation
is of ${\mathcal O}(g^4)$, the 3PI effective action provides 
a self-consistently complete starting point for its description. 
To this order the effective action resums in particular all diagrams 
enhanced by the Landau Pomeranchuk Migdal effect~\cite{Aurenche:2000gf}, 
which has been extensively discussed in recent literature in the
context of transport coefficients for gauge theories~\cite{Arnold:2002zm}.

\subsection{``On-shell'' limits}
\label{eq:onshelllimits}

The evolution equations (\ref{eq:rho})--(\ref{eq:Fexact}) to
order $g^2$ and higher contain so-called ``off-shell'' and ``memory'' 
effects due to their time integrals on the RHS. 
To simplify the description one may consider 
a number of additional assumptions which finally lead
to effective kinetic or Boltzmann-type descriptions for on-shell
particle number distributions. 
Much of this discussion is standard and can be 
found e.g.~summarized in Ref.~\cite{Blaizot:2001nr}, and we will only 
repeat what is necessary for our purposes. The derivation of kinetic equations
for the two-point functions $F^{\mu\nu}(x,y)$ and $\rho^{\mu\nu}(x,y)$ 
of Sec.~\ref{sec:frho} can be based on 
(i)~the restriction that the initial condition for the time
evolution problem is specified in the remote past,
i.e.~$t_0 \to - \infty$, (ii) a derivative expansion in the center
variable $X = (x+y)/2$, and (iii) a `quasiparticle' picture.    
To make contact with the literature we will adopt this standard 
procedure in the following and discuss limitations 
in Sec.~\ref{sec:limitations}.

For the sake of simplicity (not required), we consider the
Feynman gauge $\xi = 1$ in the following. We will also consider 
a chirally symmetric theory, i.e.~no vacuum fermion mass,
along with parity and $CP$ invariance. Therefore, the system is
charge neutral and, in particular, the most general fermion
two-point functions can be written in terms of vector components
only~\cite{Berges:2002wr}: 
$F_{\Delta}(x,y) = \gamma_{\mu} F_{\Delta}(x,y)^{\mu}$,
$\rho_{\Delta}(x,y) = \gamma_{\mu} \rho_{\Delta}(x,y)^{\mu}$,
with hermiticity properties $F_{\Delta}(x,y)^{\mu} = 
[F_{\Delta}(y,x)^{\mu}]^*$, $\rho_{\Delta}(x,y)^{\mu}
= - [\rho_{\Delta}(y,x)^{\mu}]^*$. For the gauge fields the
respective properties of the statistical and spectral
correlators read $F_D(x,y)^{\mu\nu} = 
[F_D(y,x)^{\nu\mu}]^*$, $\rho_D(x,y)^{\mu\nu}
= - [\rho_D(y,x)^{\nu\mu}]^*$. 

In order to Fourier transform with respect to the relative coordinate
$s^{\mu} = x^{\mu} - y^{\mu}$, we write
\bea
\tilde{F}_D \left( X, k \right)^{\mu\nu} &=& \int {\rm d}^4{s} \; e^{i k s} 
F_D
\left( X + \frac{s}{2}, X - \frac{s}{2} \right)^{\mu\nu} \;, 
\\
\tilde{\varrho}_D
\left( X, k \right)^{\mu\nu} &=& - i \int {\rm d}^4{s} \; e^{i k s} 
\rho_D \left( X + \frac{s}{2}, X - \frac{s}{2} \right)^{\mu\nu} \;, 
\label{eq:wigner}
\eea
and equivalently for the fermion statistical and spectral
function, $\tilde{F}_{\Delta} (X,k)$
and $\tilde{\varrho}_{\Delta} (X,k)$.
Here we have introduced a factor $-i$ in the definition of the 
spectral function transform for convenience.
For the Fourier transformed quantities we note the 
following hermiticity properties, for the
gauge fields:
$[\tilde{F}_D(X,k)^{\mu\nu}]^* = 
\tilde{F}_D(X,k)^{\nu\mu} \,\, , \,\,
[\tilde{\varrho}_D(X,k)^{\mu\nu}]^* = 
\tilde{\varrho}_D(X,k)^{\nu\mu}$,
and for the vector components of the fermion fields:
$[\tilde{F}_{\Delta}(X,k)^{\mu}]^* = 
\tilde{F}_{\Delta}(X,k)^{\mu} \,\, , \,\,
[\tilde{\varrho}_{\Delta}(X,k)^{\mu}]^* = 
\tilde{\varrho}_{\Delta}(X,k)^{\mu}$.
After sending 
$t_0 \to - \infty$ the derivative expansion can be
efficiently applied to the exact Eqs.~(\ref{eq:rho})---(\ref{eq:Fexact}). 
Here one considers the difference of (\ref{eq:rho}) and the
one with interchanged coordinates $x$ and $y$, and equivalently for
the other equations. We use 
\bea
\int\! {\rm d}^4{s} \; e^{i k s} \int\! {\rm d}^4{z}
f(x,z) g(z,y) &\!=\!& \tilde{f}(X,k) \tilde{g}(X,k) + \ldots 
\label{eq:grad1} \,\, \label{eq:grad2}\\
\int\! {\rm d}^4{s} \; e^{i k s} \int\! {\rm d}^4{z}\int\! {\rm d}^4{z'}
f(x,z) g(z,z') h(z',y) &\!=\!& \tilde{f}(X,k) \tilde{g}(X,k) 
\tilde{h}(X,k) + \ldots 
\nonumber 
\eea 
where the dots indicate derivative terms, which will be
neglected. E.g.~the first derivative corrections
to (\ref{eq:grad1}) can be written as a Poisson bracket~\cite{Blaizot:2001nr}, 
which is in particular important if `finite-width' effects of the spectral
function are taken into account. However, a typical quasiparticle picture 
which employs a free-field or `zero-width' form 
of the spectral function is consistent with
neglecting derivative terms in the scattering part.
We also note that the 
quasiparticle/free-field form of the two-point functions implies 
\bea
F_D(X,k)^{\mu\nu} \to - g^{\mu\nu} F_D(X,k)
\quad , \quad \rho_D(X,k)^{\mu\nu} \to - g^{\mu\nu} \rho_D(X,k) \, .
\label{eq:simpleindex}
\eea  
At this point the only use of the above replacement is that all 
Lorentz contractions can be done. This doesn't affect the derivative
expansion but keeps the notation simple. Similar to 
Eq.~(\ref{eq:wigner}), we define the Lorentz contracted self-energies:
\bea
- 4 \tilde{\Pi}_{(F)} (X,k) &\equiv& \int {\rm d}^4{s} \; e^{i k s} 
\Pi_{(F)}{\left( X + \frac{s}{2}, X - \frac{s}{2} \right)^{\mu}}_{\mu} \, ,
\\
-4 \tilde{\Pi}_{(\varrho)} (X,k) &\equiv& -i \int {\rm d}^4{s} \; e^{i k s} 
\Pi_{(\rho)}{\left( X + \frac{s}{2}, X - \frac{s}{2} \right)^{\mu}}_{\mu} \, . 
\eea
Without further assumptions, i.e.~using the above notation and 
applying the approximation (\ref{eq:grad1}) and 
(\ref{eq:simpleindex}) to the exact evolution equations
one has~(cf.~also~\cite{Berges:2002wt})\footnote{The relation
to a more conventional form of the equations can be seen by writing:
\bea
\lefteqn{\left(
\tilde{\Pi}_{(\varrho)} \tilde{F}_D  
- \tilde{\Pi}_{(F)} \tilde{\varrho}_D \right) (X,k) = }
\nonumber \\
&&
\left( \left[ \tilde{\Pi}_{(F)} + \frac{1}{2} \tilde{\Pi}_{(\varrho)} \right]
\left[\tilde{F}_D - \frac{1}{2} \tilde{\varrho}_D \right] -
\left[ \tilde{\Pi}_{(F)} - \frac{1}{2} \tilde{\Pi}_{(\varrho)} \right]
\left[\tilde{F}_D + \frac{1}{2} \tilde{\varrho}_D \right] 
\right) (X,k) \, .\nonumber
\eea
The difference of the
two terms on the RHS can be directly interpreted as the
difference of a so-called `loss' and a `gain' term in
a Boltzmann-type description.
} 
\bea
2\, k^{\mu} \frac{\partial}{\partial X^{\mu}} \tilde{F}_D(X,k)
&=&  \tilde{\Pi}_{(\varrho)} (X,k)\, \tilde{F}_D (X,k)  
- \tilde{\Pi}_{(F)} (X,k)\, 
\tilde{\varrho}_D (X,k) \, , \label{eq:F2} \qquad \\
2\, k^{\mu} \frac{\partial}{\partial X^{\mu}} 
\tilde{\varrho}_D (X,k) &=& 0 \, . \label{eq:rho2}
\eea 
One observes that the equations (\ref{eq:F2})
and (\ref{eq:rho2}) have a 
structure reminiscent of that for the exact equations for vanishing
`background' fields, (\ref{eq:rho}) and (\ref{eq:F}),
evaluated at equal times $x^0 = y^0$. 
However, one should keep 
in mind that (\ref{eq:F2}) and (\ref{eq:rho2}) are, in 
particular, only valid for initial
conditions specified in the remote past and neglecting gradients
in the collision part. 

From (\ref{eq:rho2}) one observes that in this approximation
the spectral function receives no contribution from scattering
described by the RHS of the exact equation (\ref{eq:rho}).
As a consequence, the spectral function obeys the
free-field equations of motion. In particular, 
$\rho_D^{\mu\nu}(x,y)$ 
have to fulfill the equal-time commutation relations 
$[\rho_D^{\mu\nu}(x,y)]_{x^0=y^0} = 0$ and
$[\partial_{x^0} \rho_D^{\mu\nu}(x,y)]_{x^0=y^0} = - g^{\mu\nu}
\delta(\bx -\by)$ in Feynman gauge. The Wigner transformed
free-field solution solving (\ref{eq:rho2}) then reads   
$\tilde{\varrho}_D (X,k) = \tilde{\varrho}_D (k) = 
 2 \pi\, {\rm sign}(k^0)\, \delta(k^2)$.
A very similar discussion can be done as well 
for the evolution equations
(\ref{eq:rhoexact}) and (\ref{eq:Fexact}) for fermions, which
is massless due to chiral symmetry as stated above.
Again, in lowest order in the derivative expansion the fermion spectral 
function obeys the free-field equations of motion and one has
$\tilde{\varrho}_{\Delta} (X,k) =
\tilde{\varrho}_{\Delta} (k) = 2 \pi k\, \slash\, {\rm sign}(k^0)\, 
\delta(k^2)$.

\subsubsection{Vanishing of the ${\mathcal O}(g^2)$ on-shell contributions}
\label{sec:onshellg2}

Assuming a ``generalized
fluctuation-dissipation relation'' or so-called
Kadanoff-Baym ansatz~\cite{KadanoffBaym}:
\bea
\tilde{F}_D (X,k) &=& \left[ \frac{1}{2} + n_D(X,k)  \right]
\tilde{\varrho}_D(X,k) \, , \nonumber\\
\tilde{F}_{\Delta} (X,k) &=& \left[ \frac{1}{2} - n_{\Delta}(X,k) \right]
\tilde{\varrho}_{\Delta} (X,k) \, ,
\label{eq:genflucdiss}
\eea
one may extract the kinetic equations for the effective
photon and fermion particle numbers $n_D$ and $n_{\Delta}$, respectively. 
Considering spatially homogeneous, isotropic systems for 
simplicity, we define the on-shell quasiparticle numbers 
($t \equiv X^0$)
\beq
n_D(t,\bk) \equiv n_D(t,k)|_{k^0 = \ak} \quad ,
\quad n_{\Delta}(t,\bk) \equiv n_{\Delta}(t,k)|_{k^0 = \ak}
\eeq
and look for the evolution equation for 
$n_D(t,\bk)=n_D(t,\ak)$. Here 
it is useful to note 
the symmetry properties
\bea
\tilde{F}_D(t,-k) &=& 
\tilde{F}_D(t,k) \,\, , \,\,
\tilde{\varrho}_D(t,-k) = 
- \tilde{\varrho}_D(t,k) \, ,
\nonumber\\
\tilde{F}_{\Delta}(t,-k)^{\mu} &=& 
- \tilde{F}_{\Delta}(t,k)^{\mu} \,\, , \,\,
\tilde{\varrho}_{\Delta}(t,-k)^{\mu} = 
 \tilde{\varrho}_{\Delta}(t,k)^{\mu} \, .
\eea
Applied to the quasiparticle ansatz (\ref{eq:genflucdiss}) these
imply
\bea
n_D(t,-k) = - \left[ n_D(t,k) + 1 \right] \,\,\, , \,\,\,
n_{\Delta}(t,-k) = - \left[ n_{\Delta}(t,k)-1 \right]
\eea
This is employed to rewrite terms with negative values of $k^0$.
To order $g^2$ the self-energies read (cf.~Eq.~(\ref{eq:selfg2})):
\bea
\tilde{\Pi}_{(F)} (X,k) &=& 2 g^2 \int \frac{{\rm d}^4 p}{(2\pi)^4}
\Big[\tilde{F}_{\Delta}(X,k+p)^{\mu}\tilde{F}_{\Delta}(X,p)_{\mu} 
\nonumber\\
&& - \frac{1}{4} \tilde{\varrho}_{\Delta}(X,k+p)^{\mu}
\tilde{\varrho}_{\Delta}(X,p)_{\mu} \Big] \, ,
\nonumber\\[0.1cm]
\tilde{\Pi}_{(\varrho)} (X,k) &=& 2 g^2 \int \frac{{\rm d}^4 p}{(2\pi)^4}
\Big[\tilde{F}_{\Delta}(X,k+p)^{\mu}\tilde{\varrho}_{\Delta}(X,p)_{\mu} 
\nonumber\\
&& - \tilde{\varrho}_{\Delta}(X,k+p)^{\mu}
\tilde{F}_{\Delta}(X,p)_{\mu} \Big] \, .
\label{eq:derivself}
\eea 
From the equations (\ref{eq:F2}) and (\ref{eq:genflucdiss})
one finds at this order: ($\bq \equiv \bk - \bp$)  
\bea
\lefteqn{
\partial_t n_D(t,\ak) = g^2 k^2 \int \frac{{\rm d}^3 p}{(2\pi)^3}\,
\frac{1}{2 \ak 2 \ap 2 \aq} \Bigg\{ }
\nn
&& \Big( n_{\Delta}(t,\ap)\, n_{\Delta}(t,\aq) \left[n_D(t,\ak) + 1\right]
\nn
&& - \left[n_{\Delta}(t,\ap)-1\right] \left[n_{\Delta}(t,\aq)-1\right] 
n_D(t,\ak) \Big)
2 \pi \delta(\ak - \ap - \aq)
\nonumber \\[0.1cm]
&+& 2 \Big( \left[n_{\Delta}(t,\ap)-1\right] n_{\Delta}(t,\aq) 
\left[n_D(t,\ak) + 1\right]
\nn
&& - n_{\Delta}(t,\ap) \left[n_{\Delta}(t,\aq)-1\right] n_D(t,\ak) \Big)
2 \pi \delta(\ak + \ap - \aq)
\nonumber \\[0.1cm]
&+& \Big( \left[n_{\Delta}(t,\ap)-1\right] \left[n_{\Delta}(t,\aq)-1\right] 
\left[n_D(t,\ak) + 1\right]
\nn
&& - n_{\Delta}(t,\ap)\, n_{\Delta}(t,\aq)\, n_D(t,\ak) \Big)
2 \pi \delta(\ak + \ap + \aq)
\Bigg\} \, .
\label{eq:nevol}
\eea
The RHS shows the standard ``gain term'' minus ``loss term'' structure.
E.g.~for the case $k^2 > 0$, $k^0 >0$ the interpretation is given by the
elementary processes $e \bar{e} \to \gamma$,
$e \to e \gamma$, $\bar{e} \to \bar{e} \gamma$ and ``$0$'' 
$\to e \bar{e} \gamma$ from which only the first one is not kinematically
forbidden. From (\ref{eq:nevol}) one also recovers the fact 
that the on-shell evolution with $k^2 = 0$ vanishes identically 
at this order. A nonvanishing result is obtained if one takes into
account off-shell corrections for a fermion line in the loop of the 
self-energy (\ref{eq:derivself}). As a consequence the first 
nonzero contribution to the self-energy starts at ${\mathcal O}(g^4)$,
which will be discussed together with the LPM enhanced
contributions below.

\subsubsection{Contributions from the self-energy to ${\mathcal O}(g^4)$}

\begin{figure}[t]
\begin{center}
\epsfig{file=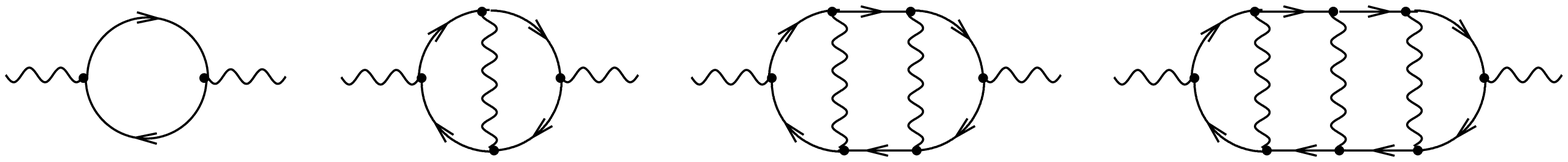,width=12.cm}
\end{center}
\vspace*{-0.5cm}
\caption{\small Infinite series of self-energy contributions with
dressed propagator lines and classical vertices.}
\label{fig:ymselfinf}
\end{figure}

It has been pointed out  
that perturbative processes in high temperature gauge theories
which are formally higher order in the weak coupling can in fact be
strongly enhanced by collinear singularities~\cite{Aurenche:2000gf}.
Recently, a kinetic description has been presented
for calculating transport coefficients in 
gauge theories at leading order in the coupling \cite{Arnold:2002zm}.
On the effective action level this can be related to considering 
an infinite series of 2PI
diagrams, and it was argued that a loop-expansion of the 2PI effective action
is not suitable in the on-shell limit~\cite{Moore:2002mt}.
For the self-energy this represents a 
series of graphs where any number of uncrossed  
lines is permitted as shown in Fig.~\ref{fig:ymselfinf}. Here propagator 
lines correspond to self-energy resummed
propagators whereas all vertices are given by the classical ones. 
We will see in the following that
the corresponding contributions to the self-energy
can be conveniently expressed using higher effective actions.

Since the lowest order contribution to the kinetic equation
is of ${\mathcal O}(g^4)$, the 3PI effective action provides 
a self-consistently complete starting point for its description. 
At this order the self-energies and vertex are given by 
Eqs.~(\ref{eq:selffermexact}), (\ref{eq:selfexact}) and (\ref{eq:vertexeom}). 
Starting from the three-vertex (\ref{eq:vertexeom}) consider first 
the vertex resummation for the photon leg only, i.e.~approximate the
fermion-photon vertex by the classical vertex. As a consequence,
one obtains:  
\bea
V^{\mu}(x,y;z) &\simeq & \gamma^{\mu} 
\delta(x - z)\delta(z - y) \label{eq:detver}\\
&& - g^2 \int_{x' y'}
\gamma^{\nu} \Delta (x,z)
V^{\mu}(x',y';z) \Delta (y',y)
\gamma^{\sigma} D_{\sigma \nu} (y,x) \, . 
\nonumber
\eea
Using this expression for the photon self-energy~(\ref{eq:selfexact}), by
iteration one observes that this resums all 
the ladder diagrams shown in Fig.~\ref{fig:ymselfinf}. 
In the context of kinetic equations,
relevant for sufficiently homogeneous systems, the dominance of this sub-class 
of diagrams has been discussed in detail in the weak coupling limit
in Ref.~\cite{Arnold:2002zm}. 
It has been suggested to decompose the contributions
to the kinetic equation into $2 \leftrightarrow 2$ particle processes,
such as $e \bar{e} \to \gamma \gamma$ annihilation in the context of QED,
and inelastic ``$1 \leftrightarrow 2$'' processes, such as the 
nearly collinear bremsstrahlung process. For the description of
``$1 \leftrightarrow 2$'' processes, 
once Fourier transformed with respect to the 
relative coordinates, the gauge field propagator in (\ref{eq:detver}) is
required for space-like momenta~\cite{Arnold:2002zm}.  
Furthermore, as indicated at 
the end of Sec.~\ref{sec:onshellg2}, the proper inclusion of nonzero 
contributions from $2 \leftrightarrow 2$ processes requires to go beyond
the naive on-shell limit. 

In the context of the evolution equations (\ref{eq:rho}) and
(\ref{eq:F}) this can be achieved by the following identities
(cf.~also Ref.~\cite{Blaizot:1999xk}):   
\bea
F_D^{\mu\nu}(x,y) &=& \lim_{t_0 \to -\infty}
\int_{t_0}^{x^0} {\rm d}z \int_{t_0}^{y^0} {\rm d}z'  
\left[ \rho_D(x,z) \Pi_{(F)}(z,z') \rho_D(z',y) \right]^{\mu\nu}
\,\, \nonumber \\
&=& - \int_{-\infty}^{\infty} {\rm d}z {\rm d}z' 
\left[ D_R(x,z) \Pi_{(F)}(z,z') D_A(z',y)\right]^{\mu\nu} \,\, 
\nonumber \\
\rho_D^{\mu\nu}(x,y) &=& \lim_{t_0 \to -\infty}
\int_{t_0}^{x^0} {\rm d}z \int_{t_0}^{y^0} {\rm d}z'  
\left[ \rho_D(x,z) \Pi_{(\rho)}(z,z') \rho_D(z',y) \right]^{\mu\nu}
\nonumber \\
&=& - \int_{-\infty}^{\infty} {\rm d}z {\rm d}z' 
\left[ D_R(x,z) \Pi_{(\rho)}(z,z') D_A(z',y)\right]^{\mu\nu} \, ,
\label{eq:Frhoident}
\eea
written in terms of the retarded and advanced propagators, 
$D_R(x,y)^{\mu\nu} = \Theta (x^0-y^0) \rho_D(x,y)^{\mu\nu}$
and $D_A(x,y)^{\mu\nu} = - \Theta (y^0-x^0) \rho_D(x,y)^{\mu\nu}$,
in order to have an unbounded time integration. The above
identity follows from a straightforward application of the
exact evolution equations and using the anti-symmetry property of
the photon spectral function, $\rho_D^{\mu\nu}(x,y)|_{x^0=y^0} = 0$. 
We emphasize that the
identity does not hold for an initial value problem
where the initial time $t_0$ is finite. Similarly, one
finds from (\ref{eq:rhoexact}) and (\ref{eq:Fexact}) 
for the fermion two-point functions using 
$\gamma^0 \rho_{\Delta}(x,y)|_{x^0=y^0} = i \delta(\bx - \by)$:
\bea
F_{\Delta}(x,y) 
&=& - \int_{-\infty}^{\infty} {\rm d}z {\rm d}z' 
\Delta_R(x,z) \Sigma_{(F)}(z,z') \Delta_A(z',y) \, , 
\nonumber \label{eq:fermidF}\\
\rho_{\Delta} (x,y) 
&=& - \int_{-\infty}^{\infty} {\rm d}z {\rm d}z' 
\Delta_R(x,z) \Sigma_{(\rho)}(z,z') \Delta_A(z',y) \,\, ,
\label{eq:fermidrho}
\eea
with $\Delta_R(x,y) = \Theta (x^0-y^0) \rho_{\Delta}(x,y)$ and
$\Delta_A(x,y) = - \Theta (y^0-x^0) \rho_{\Delta}(x,y)$.
Neglecting all derivative
terms, i.e.~using (\ref{eq:grad2}), and the above notation
these give:\footnote{As for the spectral function $\varrho (X,k)$
in Eq.~(\ref{eq:wigner}), the Fourier transform of the retarded
and advanced propagators includes a factor
of $-i$.}  
\bea
\tilde{F}_D(X,k) &\simeq& 
 \tilde{D}_R(X,k) \tilde{\Pi}_{(F)}(X,k) 
\tilde{D}_A(X,k) \, ,
\nonumber\\
\tilde{\varrho}_D (X,k) &\simeq& 
 \tilde{D}_R(X,k) \tilde{\Pi}_{(\varrho)}(X,k) 
\tilde{D}_A(X,k) \, ,
\label{eq:psident}
\eea
and equivalently for the fermion two-point functions.
Applied to one fermion line in the one-loop contribution of 
Fig.~\ref{fig:ymselfinf}, it is straightforward to recover the 
standard Boltzmann equation for $2 \leftrightarrow 2$ processes, using
the ${\mathcal O}(g^2)$ fermion self-energies:
\bea
\tilde{\Sigma}_{(F)} (X,k)^{\mu} &=& - 2 g^2 \int \frac{{\rm d}^4 p}{(2\pi)^4}
\Big[\tilde{F}_D(X,p) \tilde{F}_{\Delta}(X,k-p)^{\mu} 
\nonumber\\
&& + \frac{1}{4} \tilde{\varrho}_D(X,p)
\tilde{\varrho}_{\Delta}(X,k-p)^{\mu} \Big] \, ,
\nonumber\\[0.1cm]
\tilde{\Sigma}_{(\varrho)} (X,k)^{\mu} 
&=& - 2 g^2 \int \frac{{\rm d}^4 p}{(2\pi)^4}
\Big[\tilde{F}_D(X,p) \tilde{\varrho}_{\Delta}(X,k-p)^{\mu} 
\nonumber\\
&& + \tilde{\varrho}_D(X,p)
\tilde{F}_{\Delta}(X,k-p)^{\mu} \Big] \, .
\label{eq:derivselfferm}
\eea 
For the Boltzmann equation
$\Delta_R$ and $\Delta_A$ are taken to enter the scattering matrix element,
which is evaluated in (e.g.~HTL resummed) equilibrium, whereas
all other lines are taken to be on-shell as in Sec.~\ref{sec:onshellg2}.  
The contributions from the 
$1 \leftrightarrow 2$ processes can be efficiently obtained following the
arguments of Ref.~\cite{Arnold:2002zm} with the help
of (\ref{eq:psident}) with the ${\mathcal O}(g^2)$ photon 
self-energies~(\ref{eq:derivself}).
Of course, simply adding the contributions from $2 \leftrightarrow 2$ 
processes and $1 \leftrightarrow 2$ processes entails the problem of
double counting since a diagram enters twice.    
This occurs whenever the internal line in a $2 \leftrightarrow 2$
process is kinematically allowed to go on-shell. This does not
happen in equilibrium and can be suppressed for the 
cases of interest~\cite{Arnold:2002zm}. 

\subsection{Discussion}
\label{sec:limitations}

In view of the generalized fluctuation-dissipation relation
(\ref{eq:genflucdiss}) employed in the above ``derivation'',
one could be tempted to say that for consistency 
an equivalent relation should be valid for the self-energies as well:
\bea
\tilde{\Pi}_{(F)} (X,k) &=& \left[ \frac{1}{2} + n_D(X,k)  \right]
\tilde{\Pi}_{(\varrho)}(X,k) \, .
\label{eq:flucdissself}
\eea
Such a relation is indeed valid in thermal equilibrium, where all 
dependence on the center coordinate $X$ is lost. 
Furthermore, the above relation can be shown to be a consequence
of (\ref{eq:genflucdiss}) using the identities (\ref{eq:Frhoident}) 
in a lowest-order derivative expansion: Together with Eq.~(\ref{eq:psident})
the above relation for the self-energies is a direct consequence of
the ansatz (\ref{eq:genflucdiss}). However, clearly
this is too strong a constraint since the evolution equation 
(\ref{eq:F2}) would become trivial in this case: Eq.~(\ref{eq:genflucdiss})
and (\ref{eq:flucdissself}) lead to a vanishing
RHS of the evolution equation for $\tilde{F}_D (X,k)$ and 
there would be no evolution.

The above argument is just a manifestation of the well-known
fact that the kinetic equation is not a self-consistent approximation
to the dynamics. The discussion of 
Sec.~\ref{eq:onshelllimits} takes 
into account the effect of scattering for the dynamics of effective 
occupation numbers, while keeping the spectrum   
free-field theory like. In contrast, the same scattering does 
induce a finite width for the spectral function in the 
self-consistent approximation discussed in Sec.~\ref{sec:frho} 
because of a nonvanishing imaginary part of the self-energy
(cf.~also the discussion and explicit solution of a similar 
Yukawa model in Ref.~\cite{Berges:2002wr}). 

Though particle number is not well-defined in an 
interacting relativistic quantum field theory in the
absence of conserved charges,
the concept of time-evolving effective particle numbers in an
interacting theory is useful in the
presence of a clear separation of scales.
Much progress has been achieved in the quantitative 
understanding of kinetic descriptions in the vicinity of thermal 
equilibrium for gauge theories at high temperature, which is well 
documented in the
literature\footnote{For recent discussions to go beyond near-equilibrium
see also Ref.~\cite{Arnold:2002zm}}~(see 
e.g.~Ref.~\cite{Blaizot:1999ip,Arnold:2002zm} 
and references therein).

A derivative expansion is typically not valid 
at early times where the time evolution can
exhibit a strong dependence on $X$,
and the homogeneity requirement underlying kinetic descriptions
may only be fulfilled at sufficiently late times. This has
been extensively discussed in the context of 
scalar~\cite{Berges:2001fi,Berges:2000ur,Aarts:2001qa}
or fermionic theories~\cite{Berges:2002wr}. 
Homogeneity is certainly realized at late times sufficiently close to 
the thermal limit, since for thermal equilibrium the correlators do 
strictly not depend on $X$. 
Of course, by construction kinetic equations 
are not meant to discuss the detailed early-time
behavior since the initial time $t_0$
is sent to the remote past. For practical purposes, in this context 
one typically specifies the initial condition for the effective 
particle number distribution at some finite
time and approximates the evolution by the 
equations with $t_0 \to - \infty$. 
The role of finite-time effects has been controversially discussed
in the recent literature in the context of photon production
in relativistic plasmas at high temperature~\cite{Fraga:2003sn}. 
Here a solution
of the proper initial-time equations as discussed
in Sec.~\ref{sec:nonequilibrium} seems mandatory.

\section{Conclusions}

Self-consistently complete loop or coupling expansions
of $n$PI effective actions are promising candidates for a 
uniquely suitable description of both nonequilibrium as well 
as equilibrium (or vacuum) quantum field theory.
It is interesting to observe that the need for a description 
of a universal late-time behavior and thermalization 
leads already for weakly-coupled quantum field theories 
to similar techniques than those employed in equilibrium
strong interaction physics.
For gauge theories, so far their use is maybe best understood
for a ``derivation'' of kinetic equations in the presence of a
weak coupling at high temperature. 
Here the employed on-shell limit circumvents problems of
gauge invariance or subtle aspects of renormalization.
Recently, a first successful implementation of a renormalization
prescription for 2PI effective actions in scalar field 
theories has been presented~\cite{Blaizot:2003an,vanHees:2001ik}. 
A prescription for gauge
theories along these lines has not been given so far 
and will be investigated in a separate work~\cite{BBRS}.
A successful completion of this program would give  
the striking prospect to solve initial-value problems in realistic
quantum field theories relevant for heavy-ion collisions.\\[0.6cm]

\noindent
{\bf \large Acknowledgements}
\vspace*{0.3cm}

\noindent
I would like to thank H.~Gies and C.S.~Fischer for interesting discussions,
and Sz.~Bors{\'a}nyi, M.M.~M{\"u}ller and J.~Serreau for collaboration on
related work.

\appendix

\section{}

We use the short-hand notation
\beq
\Theta(x^0,y^0,z^0) \equiv \Theta(x^0-y^0)\Theta(y^0-z^0) \, .
\eeq
With the separation of Eq.~(\ref{eq:separation}), the  
time-ordered three-vertex can be written as
\bea
\bar{V}^{\mu} (x,y;z) &=& V_{(1)}^{\mu} (x,y;z) \Theta(x^0,y^0,z^0)
+ V_{(2)}^{\mu} (x,y;z) \Theta(y^0,z^0,x^0)
\nonumber\\
&+& V_{(3)}^{\mu} (x,y;z) \Theta(z^0,x^0,y^0)
+ V_{(4)}^{\mu} (x,y;z) \Theta(z^0,y^0,x^0) \qquad
\label{eq:vstart}\\
&+& V_{(5)}^{\mu} (x,y;z) \Theta(x^0,z^0,y^0)
+ V_{(6)}^{\mu} (x,y;z) \Theta(y^0,x^0,z^0) \, ,
\nonumber
\eea
with `coefficients' $V_{(i)}^{\mu} (x,y;z)$, $i = 1,\ldots, 6$. 
These coefficients can be expressed in terms of three spectral vertex
functions $U_{(\rho)}^{\mu}(x,y;z)$, $V_{(\rho)}^{\mu}(x,y;z)$ and 
$W_{(\rho)}^{\mu}(x,y;z)$, as well as the corresponding statistical components 
$U_{(F)}^{\mu}(x,y;z)$, $V_{(F)}^{\mu}(x,y;z)$ and $W_{(F)}^{\mu}(x,y;z)$
that have been employed in Eq.~(\ref{eq:vbar}). One finds,
suppressing the space-time arguments:
\bea
V_{(1)}^{\mu} &\equiv& U_{(F)}^{\mu} + V_{(F)}^{\mu} - W_{(F)}^{\mu}
- \frac{i}{2} \Big( U_{(\rho)}^{\mu} + V_{(\rho)}^{\mu} 
- W_{(\rho)}^{\mu} \Big) \, ,
\nonumber\\
V_{(2)}^{\mu} &\equiv& U_{(F)}^{\mu} - V_{(F)}^{\mu} + W_{(F)}^{\mu}
- \frac{i}{2} \Big( U_{(\rho)}^{\mu} - V_{(\rho)}^{\mu} 
+ W_{(\rho)}^{\mu} \Big) \, ,
\nonumber\\
V_{(3)}^{\mu} &\equiv& - U_{(F)}^{\mu} + V_{(F)}^{\mu} + W_{(F)}^{\mu}
- \frac{i}{2} \Big( - U_{(\rho)}^{\mu} + V_{(\rho)}^{\mu} 
+ W_{(\rho)}^{\mu} \Big) \, ,
\nonumber\\
V_{(4)}^{\mu} &\equiv& U_{(F)}^{\mu} + V_{(F)}^{\mu} - W_{(F)}^{\mu}
+ \frac{i}{2} \Big( U_{(\rho)}^{\mu} + V_{(\rho)}^{\mu} 
- W_{(\rho)}^{\mu} \Big) \, ,
\\
V_{(5)}^{\mu} &\equiv& U_{(F)}^{\mu} - V_{(F)}^{\mu} + W_{(F)}^{\mu}
+ \frac{i}{2} \Big( U_{(\rho)}^{\mu} - V_{(\rho)}^{\mu} 
+ W_{(\rho)}^{\mu} \Big) \, ,
\nonumber\\
V_{(6)}^{\mu} &\equiv& - U_{(F)}^{\mu} + V_{(F)}^{\mu} + W_{(F)}^{\mu}
+ \frac{i}{2} \Big( - U_{(\rho)}^{\mu} + V_{(\rho)}^{\mu} 
+ W_{(\rho)}^{\mu} \Big) \, .
\nonumber
\eea
In terms of the coefficients $V_{(i)}^{\mu}$ these are given by:
\bea
U_{(F)}^{\mu} &=& \frac{1}{4} \Big(V_{(1)}^{\mu} 
+ V_{(2)}^{\mu} + V_{(4)}^{\mu} + V_{(5)}^{\mu}
\Big)\, , \quad 
U_{(\rho)}^{\mu} = \frac{i}{2} \Big(V_{(1)}^{\mu} 
+ V_{(2)}^{\mu} - V_{(4)}^{\mu} - V_{(5)}^{\mu}
\Big)\, , \nonumber \\
V_{(F)}^{\mu} &=& \frac{1}{4} \Big(V_{(1)}^{\mu} 
+ V_{(3)}^{\mu} + V_{(4)}^{\mu} + V_{(6)}^{\mu}
\Big)\, , \quad
V_{(\rho)}^{\mu} = \frac{i}{2} \Big(V_{(1)}^{\mu} 
+ V_{(3)}^{\mu} - V_{(4)}^{\mu} - V_{(6)}^{\mu}
\Big)\, , \quad \nonumber \\
W_{(F)}^{\mu} &=& \frac{1}{4} \Big(V_{(2)}^{\mu} 
+ V_{(3)}^{\mu} + V_{(5)}^{\mu} + V_{(6)}^{\mu}
\Big)\, , \quad
W_{(\rho)}^{\mu} = \frac{i}{2} \Big(V_{(2)}^{\mu} 
+ V_{(3)}^{\mu} - V_{(5)}^{\mu} - V_{(6)}^{\mu}
\Big) \nonumber \, .
\eea
Insertion shows the equivalence of (\ref{eq:vstart}) and
(\ref{eq:vbar}).

\end{document}